%% file: main.tex
\documentclass[twocolumn]{aastex701}

\usepackage{amsmath,amssymb,mathrsfs}
\usepackage{tabularx}
\usepackage{booktabs}
\usepackage{CJK}
\usepackage{enumerate}

\newcommand{\nusupp}{{\nu_{\mathrm{supp}}}}

\newcommand{\numberofsystems}{17 }

\begin{document}
\begin{CJK*}{UTF8}{gbsn}

\shorttitle{SLF Non-Suppression}
\shortauthors{Ma, Pedersen \& Rui}

\title{Absence of Magnetic Suppression in Massive Star Variability}

\author[orcid=0000-0001-6117-5750]{Linhao Ma (马林昊)}
\affiliation{Kavli Institute for Theoretical Physics, University of California, Santa Barbara, CA 93106, USA}
\affiliation{Department of Astrophysical Sciences, Princeton University, 4 Ivy Lane, Princeton, NJ 08544, USA}
\email{linhaoma@princeton.edu}

\author[orcid=0000-0002-7950-0061]{May G. Pedersen}
\affiliation{Sydney Institute for Astronomy, School of Physics, University of Sydney NSW 2006, Australia}
\email{may.pedersen@sydney.edu.au}

\author[orcid=0000-0002-1884-3992]{Nicholas Z. Rui}
\altaffiliation{NASA Hubble Fellow}
\affiliation{Department of Astrophysical Sciences, Princeton University, 4 Ivy Lane, Princeton, NJ 08544, USA}
\affiliation{Center for Interdisciplinary Exploration and Research in Astrophysics (CIERA), Northwestern University, 1800 Sherman Ave., Evanston, IL 60201, USA}
\email{nrui@northwestern.edu}

\begin{abstract}

Although stochastic low-frequency (SLF) variability is ubiquitous among hot, massive stars, its physical origin remains uncertain.
One proposed explanation is that this variability is produced by internal gravity waves (IGWs) excited in the stellar interior by convection in the stellar core or subsurface layers.
Strong magnetic fields, however, are expected to significantly attenuate such waves.
In this work, we model the conditions for magnetic gravity-wave suppression in \numberofsystems magnetic hot stars that exhibit significant SLF variability.
We find that in many cases, IGWs excited at frequencies comparable to the observed variability should be strongly suppressed by the magnetic field, sometimes across a substantial fraction of the frequency range containing the observed SLF power.
This finding challenges IGWs as a universal explanation for SLF variability in massive stars and motivates the consideration of alternative origins.

\end{abstract}


\section{Introduction} 

\begin{figure*}
    \centering
    \includegraphics[width=\textwidth]{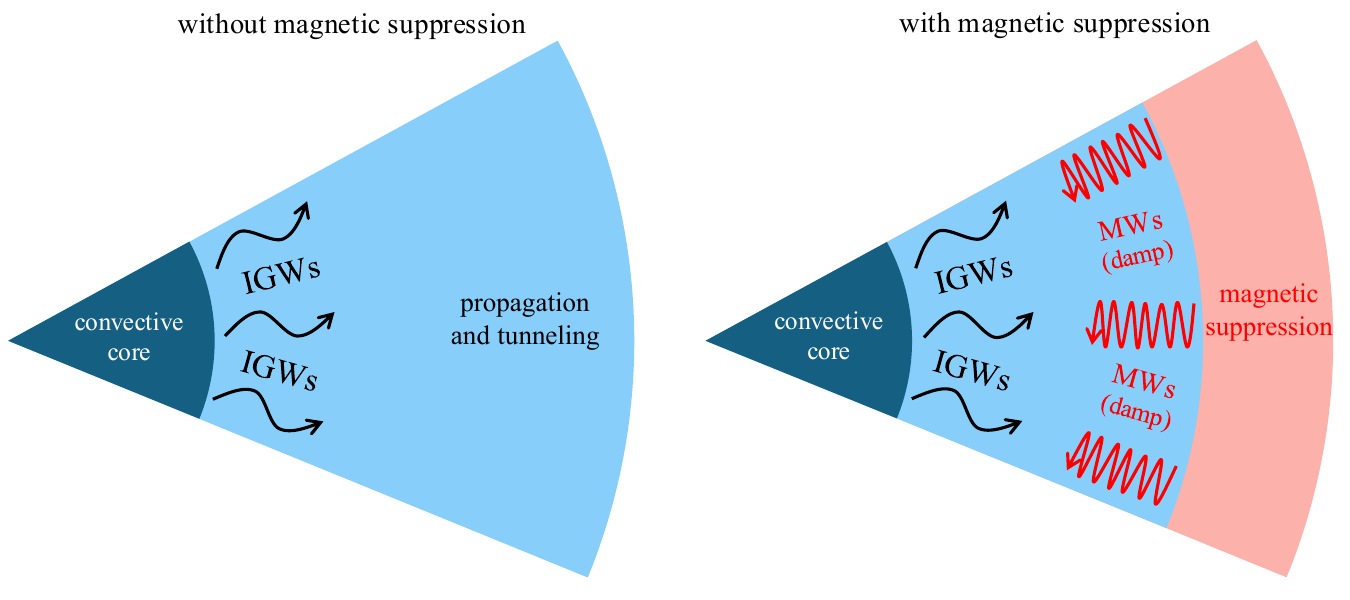}
    \caption{Schematic of magnetic suppression of internal gravity waves (IGWs) in massive stars.
    \textbf{Left}: Without magnetic suppression and in the absence of other damping, IGWs can propagate through the radiative envelope and reach the stellar photosphere, appearing as SLF variability. \textbf{Right}: In stars with strong magnetic fields, the propagation of low-frequency IGWs is suppressed in the outer envelope as the density drops.
    The outward propagating gravity waves are converted into inward propagating magnetic waves (MWs), whose short wavelengths allow them to damp efficiently.
    Such magnetically suppressed gravity waves fail to reach the surface and cannot reproduce the observed SLF phenomenon.
    }
    \label{fig:cartoon}
\end{figure*}

High-precision space-based photometry has revealed stochastic low-frequency (SLF) variability to be a common feature of hot, massive stars.
In the frequency domain, this variability appears as a broad, red-noise-like excess of power toward low frequencies, rather than sharp power spectral peaks at discrete mode frequencies.
SLF variability has now been detected across a wide range of masses and evolutionary stages among O- and B-type stars \citep[e.g.,][]{Bowman2019,Bowman2019b,Pedersen2019,Bowman2020,Ma2024,Pedersen2025,Kourniotis2025}.
Despite the ubiquity of this phenomenon, the underlying mechanism remains debated.

One common proposal interprets SLF variability as internal gravity waves (IGWs) excited inside the star.
In massive stars, these waves can be driven by turbulent motions in either their convective cores  \citep{Aerts2015,Bowman2019b,Edelmann2019} or their subsurface convection zones \citep{Pathak2026}.
The IGWs are expected to appear as SLF-like variability, provided they can propagate all the way to the surface before damping out \citep{Aerts2009,Bowman2019b}. If a substantial fraction of the observed SLF power is indeed produced by such waves, it would provide a rare observational window into otherwise inaccessible regions of massive stars. Moreover, IGWs can transport angular momentum \citep{Rogers2013, Fuller2014} and chemical species \citep{Rogers2017,Varghese2025} through stellar interiors, such that observational constraints on their properties would have important implications for models of stellar evolution.

However, IGWs are particularly susceptible to magnetic effects because their low frequencies efficiently couple them to magnetic tension.
This magnetic suppression mechanism has been invoked to explain the suppression of dipole mixed modes in red giants \citep{Fuller:2015:SuppressedDipole,Stello2016}, and it has subsequently been used to constrain internal stellar magnetic fields from gravity-mode pulsations in other stars \citep{Lecoanet2022,Rui2025,DuerfeldtPedros2026,Rui:2026:sdbmagnet}.
If SLF variability is indeed caused by IGWs, it should be absent in stars whose radiative envelopes contain strong magnetic fields.
This makes magnetic hot stars a promising laboratory for testing the hypothesis that SLF variability indeed arises from IGWs.

In this work, we investigate the conditions under which low-frequency IGWs are expected to undergo magnetic suppression in a sample of \numberofsystems magnetic hot stars.
Even under conservative assumptions, we find that the magnetic fields observed in many SLF variables are strong enough to suppress IGW propagation over a significant fraction of the frequency range occupied by the SLF signal.
Despite this, substantial SLF power is observed in precisely these frequency ranges.
Our results are in significant tension with an IGW origin for SLF variability, suggesting a different underlying physical mechanism.

This paper is organized as follows: in Section \ref{sec:theory}, we review the physical mechanism of magnetic suppression of IGWs. We describe our observations and modeling of a sample of magnetic SLF variables in Section \ref{sec:observation_and_modeling}, and analyze the magnetic suppression hypothesis in Section \ref{sec:analysis_and_results}. We find that, in many cases, IGWs should be damped at frequencies where the SLF power excess is observed.
We consider two theoretical complications in Section \ref{sec:discussions}, though we find that accounting for them does not qualitatively change our conclusions.
We conclude and discuss possible future work in Section \ref{sec:conclusion}.
Our analyses for the full sample are presented in Appendix \ref{app:full}.

\section{Theoretical background}
\label{sec:theory}

Stars more massive than $\simeq1.2\,M_\odot$ possess radiative envelopes capable of supporting buoyant motions.
The stable stratification in these envelopes can support the propagation of IGWs, potentially carrying wave energy to the photosphere where it may be observed (left panel of Figure \ref{fig:cartoon}).
If wave damping is sufficiently weak, IGWs can form high-order standing gravity (g) modes with a dense spectrum at low frequencies.
This dense forest of g modes has been proposed as the origin of observed SLF variability \citep{Bowman2019,Bowman2019b,Bowman2020}.

However, low-frequency IGWs experience magnetic suppression in the presence of sufficiently strong radial magnetic fields due to their primarily horizontal motions.
Magnetic suppression occurs roughly when the Alfv\'en frequency is at least comparable to the IGW frequency $\omega$.
This criterion can be shown \citep{Fuller:2015:SuppressedDipole,Cantiello:2016:MagnetoAstero} to imply that suppression occurs when:
\begin{equation} \label{eqn:Bcrit}
    B_{r} > B_{r,\mathrm{crit}} \simeq \frac{1}{\sqrt{\ell(\ell+1)}}\frac{\sqrt{\pi\rho}\,\omega^2r}{N}\mathrm{,}
\end{equation}
where $\rho$ is the density, $N$ is the Brunt--V\"ais\"al\"a frequency \citep{Vaisala1925}, and $\ell$ is the spherical degree of the gravity waves.
Equivalently, a radial magnetic field with a strength of $B_r$ inhibits the propagation of IGWs with frequencies below a critical frequency $\omega_B$ given by
\begin{equation}
\label{eq:omega_B}
    \omega_B = \left(\sqrt{\ell(\ell+1)}\frac{NB_r}{\sqrt{\pi\rho}r}\right)^{1/2}\,.
\end{equation}
Physically, this damping is believed to arise from a combination of conversion to easily damped slow magnetic waves and Alfv\'en waves \citep[referred to in the right panel of Figure \ref{fig:cartoon} as magnetic waves, or MWs;][]{Lecoanet2017,Lecoanet2022,Loi:2017:AlfvenResonances,Loi:2018:MagneticDynamicalChaos,Rui2023,David:2026:MagneticConversion}.

In realistic stars, $\rho$ drops sharply near the stellar surface, producing a rapid increase in $\omega_B$ there.
IGWs are especially easily 
damped in this ``magnetic suppression'' region in the outer layers of the envelope (Figure \ref{fig:cartoon}, right panel). This implies:
\begin{enumerate}
    \item Low-frequency IGWs with $\omega<\omega_B$ that are driven in the stellar interior cannot reach the photosphere with significant power.
    \item Therefore, surface variability with frequencies $\omega<\omega_B$ cannot originate from IGWs excited within the stellar interior.
\end{enumerate}

Hence, comparing the observed frequencies of SLF variability with $\omega_B$ can help test the hypothesis that SLF variability arises from internally driven IGWs.

\section{Observations and Modeling}
\label{sec:observation_and_modeling}

We obtain the light curves showing SLF variability for a sample of known magnetic hot stars from the Transiting Exoplanet Survey Satellite (\textit{TESS}, \citealt{TESS}; Section \ref{sec:observation}). We then model the critical frequency $\omega_B$ as a function of $r$ in this sample (Section \ref{sec:model}) to compare it with the frequencies of SLF variability observed.

\subsection{Sample selection and TESS observations}
\label{sec:observation}

Starting from a sample of 114 magnetic stars previously found by \citet{Shen2023} to show SLF variability, we select 49 stars from this sample identified by cross-matching the 114 stars with the polar magnetic field measurements summarized in \citet{Petit2013} and \citet{Shultz2020}. The magnetic field constraints represent the field strength extracted from spectropolarimetry/photopolarimetry under the assumption of a dipolar field topology. Since the analysis by \citet{Shen2023}, additional \textit{TESS} data have become available. We thus extract our own \textit{TESS} light curves, including the newly available data.

We extract light curves from the \textit{TESS} Full Frame Images (FFI) data for sectors $1$--$107$ (up to August 2026)\footnote{Due to changes in the observing strategy of the \textit{TESS} mission, the observing cadence of the FFI time series data has reduced over time starting from 30-min (sectors 1-26), then reduced to 10-min (sectors 27-55), and finally 200-sec (as of sector 56).} using the \texttt{tesscut} \citep{tesscut2019} and \texttt{lightkurve} \citep{Lightkurve2018} \texttt{python} packages. For each star we define an initial target aperture mask by selecting all pixels around the target where the median flux exceeds three times the standard deviation of the median flux $\sigma_{{\rm std}} \left(F_{\rm median}\right)$ over all pixels in the $20\times 20$ pixel FFI cutout centered on the target. This initial target aperture mask is then manually modified on a star-by-star basis to reduce contamination from nearby sources when possible. While inspecting the pixel data, we excluded 8 stars from the sample where contamination from nearby stars made it impossible to determine the source of the SLF variability. 

The target flux is extracted from the final target aperture mask and then background corrected. To perform the background correction, we first identify a background aperture mask containing all pixels where the median flux is less than $0.001\sigma_{{\rm std}} \left(F_{\rm median}\right)$. We carry out a principal component analysis of the background pixels and reduce the background variability due to scattered light to its first one to seven principal components, depending on the target. We then use the \texttt{RegressionCorrector} class of \texttt{lightkurve} to correct the target light curve for the background variability. The final extracted \textit{TESS} light curves are normalized by dividing each sector by a 0th--3rd order polynomial fit. All data points with \textit{TESS} quality flags different from zero are removed from the light curves. 

We perform iterative pre-whitening on the normalized light curves to remove non-SLF variability caused by, e.g., rotational modulation, binary eclipses, or stellar pulsations. We describe the details of the pre-whitening process in Appendix \ref{app:pre-whitening}. Our final sample consists of \numberofsystems SLF variables after excluding 24 targets whose light curves are dominated by instrumental noise, or other kind of variabilities. Table~\ref{tab:observed} summarizes their \textit{TESS} observing sectors and the measured values or lower limits on the polar magnetic fields. The field measurements all come from \cite{Petit2013} in the final sample.

\subsection{Determining $\omega_B$ from stellar models}
\label{sec:model}

\begin{figure*}
    \centering
    \includegraphics[width=\textwidth]{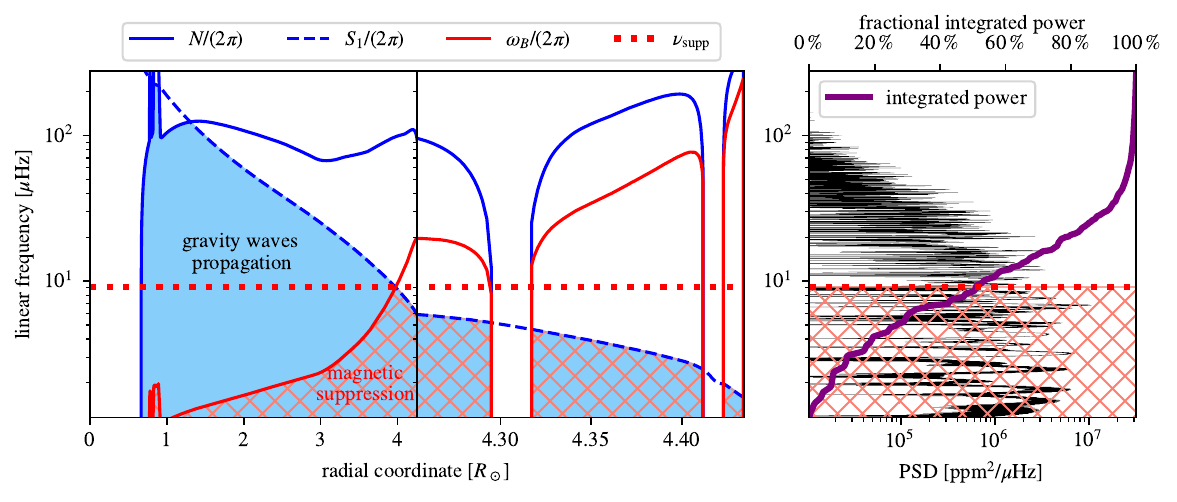}
    \caption{\textbf{Left}: Asteroseismic propagation diagram for a $7.25\,M_\odot$ main-sequence model corresponding to the massive magnetic star HD 36485, which has a measured polar magnetic field $B_\mathrm{p}\approx10\,\mathrm{kG}$.
    IGWs can propagate in the absence of a magnetic field in the gravity-mode region (blue shaded region) but are strongly magnetically damped when $\omega<\omega_B$ (red hatched region).
    The red dotted line indicates $\nusupp$, the linear frequency below which an IGW is magnetically suppressed before reaching the photosphere.
    \textbf{Right:} The observed power spectral density (PSD) for HD 36485 (black lines) and the total fractional integrated power (thick purple line) in the SLF frequency range. Roughly $51\%$ of power is below the magnetic suppression frequency, and were expected to be suppressed if they originated from IGWs that propagated from the stellar interior.
    The observed SLF variability in this star appears inconsistent with IGWs. 
    }
    \label{fig:propagation}
\end{figure*}

As $\omega_B$ depends on the internal stellar profile, we run a grid of massive stellar models with the \texttt{MESA} stellar evolution code (r24.08.1; \citealt{Paxton2011,Paxton2013,Paxton2015,Paxton2018,Paxton2019,Jermyn2023}) to create physically realistic profiles for the stars in our sample.
Our grid spans initial (zero-age main-sequence; ZAMS) masses of $5M_\odot \leq M_{\mathrm{ZAMS}} \leq 60M_\odot$ at fixed solar metallicity ($Z=0.0142$). We adopt a mass step size of $0.25\,M_\odot$ between $5$ and $10\,M_\odot$, and $0.5\,M_\odot$ between $10$ and $20\,M_\odot$. We use the same controls and setups for our stellar models as in the \texttt{MESA} test suite simulation \texttt{20M\_pre\_ms\_to\_core\_collapse}, which has been tested to produce reasonable stellar tracks on the HR diagram up to $80\,M_\odot$. All models are run from the pre-main sequence until core hydrogen depletion, when the central hydrogen mass fraction drops below $10^{-4}$.

As discussed in Section \ref{sec:theory}, magnetic suppression is most sensitive to the near-surface structure where the density is lowest. We therefore identify the best-fitting model for each star in our sample by minimizing:
\begin{equation}
\begin{split}
    \chi^2 &= \left[(\log L - (\log L)^{\mathrm{model}})/\sigma_{\log L}\right]^2 \\
    &+ \left[(\log T_{\mathrm{eff}} - (\log T_{\mathrm{eff}})^{\mathrm{model}})/\sigma_{\log T_{\mathrm{eff}}}\right]^2 \\
    &+ \left[(\log g - (\log g)^{\mathrm{model}})/\sigma_{\log g}\right]^2\,,
\end{split}
\end{equation}
where $L$, $T_\mathrm{eff}$, and $g$ are the luminosity, effective temperature, and surface gravity of the star, respectively, taken from \cite{Shen2023} and summarized in Table \ref{tab:observed}. Including $\log g$ in the fit provides an additional constraint on the stellar mass and radius, and hence on the near-surface structure most relevant for magnetic suppression. Table \ref{tab:results} summarizes the model parameters for these best-fitting models.

As our 1D stellar evolutionary models do not explicitly model magnetic fields, we assume that the internal radial field is at least as strong as the polar (dipole) field measured for these stars, i.e., $B_r=B_\mathrm{p}$ (or $B_r=B_\mathrm{p,\mathrm{min}}$ if only a lower limit of $B_\mathrm{p}$ is given).
This is a fairly conservative assumption, which is only violated if the near-surface magnetic field is extremely localized to the outermost layers of the star. For the hot stars in our sample with radiative envelopes, we expect the magnetic field to take the form of large-scale fields frozen into the stellar interior. The only scenario where the field may indeed be rooted shallowly in the envelope is when it is actively replenished by a near-surface dynamo. This likely produces significantly higher-order multi-polar field components that are not found in the 17 stars in our sample (cf. Table 1 in \citealt{Petit2013}).

Table \ref{tab:results} summarizes the $B_r$ assumed for our sample. Equation \ref{eq:omega_B} then gives $\omega_B$ based on the stellar profile of the best-fitted model and assumed $B_r$.

\section{Analysis and Results}
\label{sec:analysis_and_results}

\begin{figure}
    \centering
    \includegraphics[width=\columnwidth]{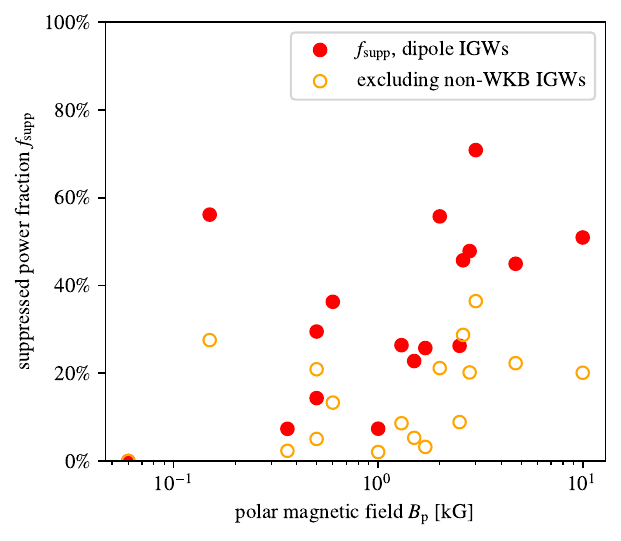}
    \caption{The expected suppressed power fraction and the strength of polar magnetic fields for the \numberofsystems magnetic stars in our analysis. The red dots show the estimates for dipole ($\ell=1$) IGWs, and the orange dots show the more conservative estimates excluding non-WKB waves. Overall, a substantial fraction of power are expected to be suppressed if they are from IGWs for many of the systems in our analysis. This questions the IGW interpretation as a common origin for stochastic low-frequency (SLF) variability.
    }
    \label{fig:suppressed_fractions}
\end{figure}

As an illustrative example, the left panel of Figure \ref{fig:propagation} shows an asteroseismic propagation diagram of the best-fitting stellar model for a representative system in our study, HD 36485 \citep[cf. Figure 3 of ][]{Fuller:2015:SuppressedDipole}.
In the absence of magnetism, dipole ($\ell=1$) IGWs propagate in regions with $\omega<\min(N,S_1)$ corresponding to the blue shaded regions in Figure \ref{fig:propagation}, where $S_1$ is the dipole Lamb frequency.
This gravity-wave region spans most of the radiative envelope, as well as a few stably stratified cavities between the subsurface convective layers (where $N$ becomes imaginary).

However, under a magnetic field with strength $B_r$, IGWs with $\omega<\omega_B$ are suppressed (the crosshatched regions in Figure \ref{fig:propagation}).
This eliminates the stably stratified propagation cavities between subsurface convective layers entirely, as $\omega_B$ is greater than $S_1$ in those regions. In the inner radiative envelope, IGWs are strongly damped once they reach the magnetic suppression region (crosshatched). This prevents them from reaching the stellar photosphere below a (linear) magnetic suppression frequency $\nusupp$, given by the local $\omega_B/2\pi$ at the crossover radius between $S_1$ and $\omega_B$. The model for HD 36485 gives $\nusupp\approx9.1\,\mathrm{\mu Hz}$.

Despite this, a substantial fraction of SLF power is indeed observed at frequencies below $\nusupp$. The right panel of Figure \ref{fig:propagation} shows the observed SLF power spectral density (PSD) of HD 36485. The purple line shows the cumulative fraction of the integrated power of the PSD \citep{Pedersen2025}:
\begin{equation}
p(\nu)\equiv\frac{\int_{\nu_\mathrm{min}}^{\nu}\mathrm{PSD}(\nu')\,d\nu'}{\int_{\nu_\mathrm{min}}^{\nu_\mathrm{norm}}\mathrm{PSD}(\nu')\,d\nu'}\,,
\end{equation}
where $\nu_\mathrm{min}=1.157\,\mu\mathrm{Hz}$, below which the power spectrum may be subject to artificial de-trending effects, and $\nu_\mathrm{norm}=278\,\mu\mathrm{Hz}$, which is the Nyquist frequency for 30-min cadence. We see that a large fraction $f_{\mathrm{supp}}\approx51\%$ of the observed SLF excess lies below $\nusupp$. HD 36485 is thus a clear-cut example of an SLF variable whose variability cannot arise from internally driven IGWs.

HD 36485 is not alone: Figure \ref{fig:suppressed_fractions} and Table \ref{tab:results} show that many of the \numberofsystems SLF variables in our sample exhibit fractions of their variability power below $\nusupp$ (Appendix \ref{app:full} presents our analysis for each star individually). As shown in Equation \ref{eq:omega_B}, IGWs with a higher spherical degree $\ell$ are subject to magnetic suppression at even higher frequencies, and a similar analysis for quadrupole ($\ell=2$) IGWs reaches the same conclusion, which is also summarized in Table \ref{tab:results}.
These examples make it clear that the IGW hypothesis for SLF variability is not viable in these stars.

\section{Discussions}
\label{sec:discussions}

\subsection{Residual Wave Power from Evanescent Tunneling}

\begin{figure*}
    \centering
    \includegraphics[width=\textwidth]{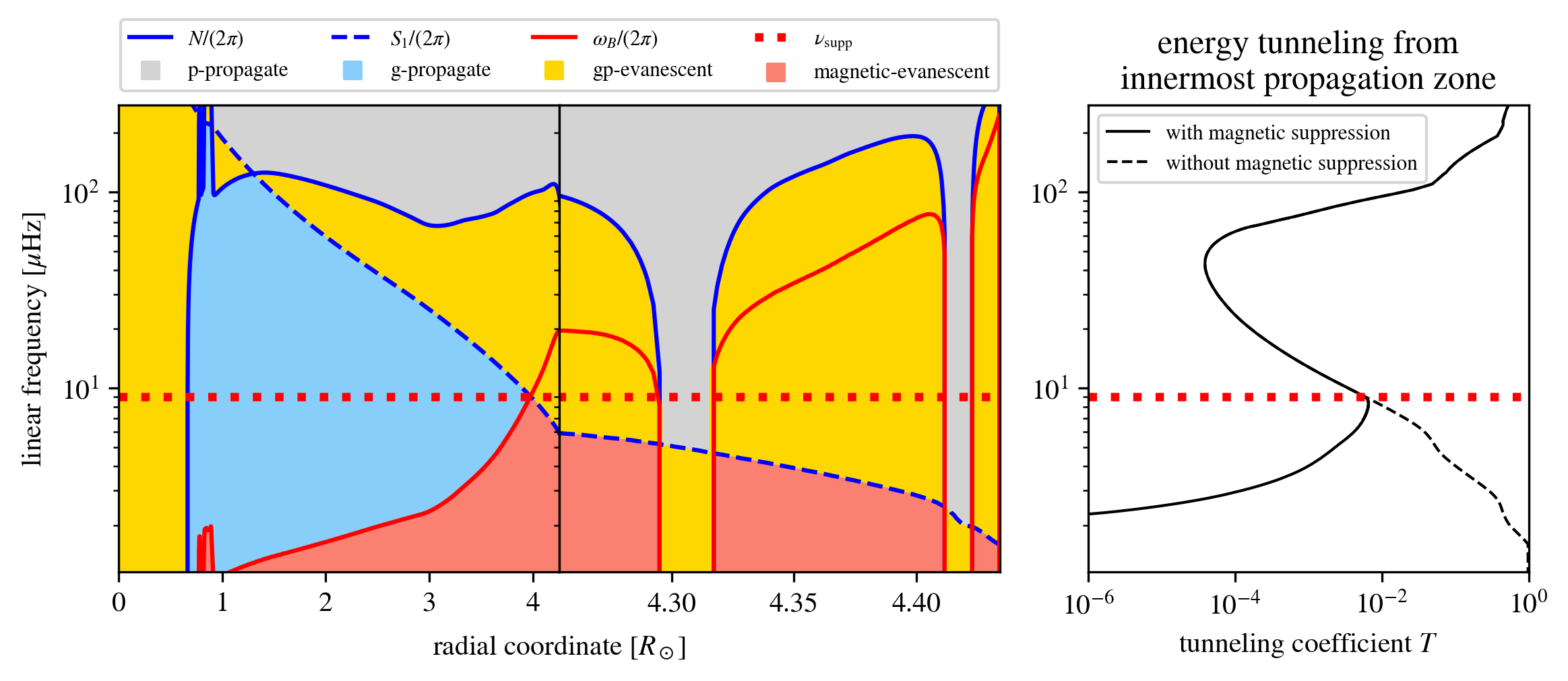}
    \caption{\textbf{Left}: Propagating and evanescent regions for internal waves for the HD 36485 model in Figure \ref{fig:propagation}.
    The blue and gray regions support the propagation of gravity and pressure waves.
    In the gold regions, waves are evanescent according to the standard non-magnetic gravito-acoustic (gp) dispersion relation and are attenuated by a factor $\exp(-2\int |\Im(k_{r,\mathrm{gp}})|\,dr)$, where $k_{r,\mathrm{gp}}$ is given by Equation \ref{eq:kr_gp}.
    In the red regions, the waves are also evanescent, but due to magnetic suppression. The attenuation factor is instead given by $\exp(-2\int |\Im(k_{r,\mathrm{mg}})|\,dr)$, where $k_{r,\mathrm{mg}}$ is given by Equation \ref{eq:kr_mg}.
    \textbf{Right}: The wave energy tunneling coefficient $T$ (solid black line) representing residual wave energy which survives evanescent decay from the propagating regions to the stellar surface.
    The small values of $T$ compared to the case without magnetic effects (dashed black line) validate our assumption that magneto-gravity waves damp out immediately upon reaching the magnetic suppression region.
    }
    \label{fig:tunneling}
\end{figure*}

\begin{figure*}
    \centering
\includegraphics[width=\textwidth]{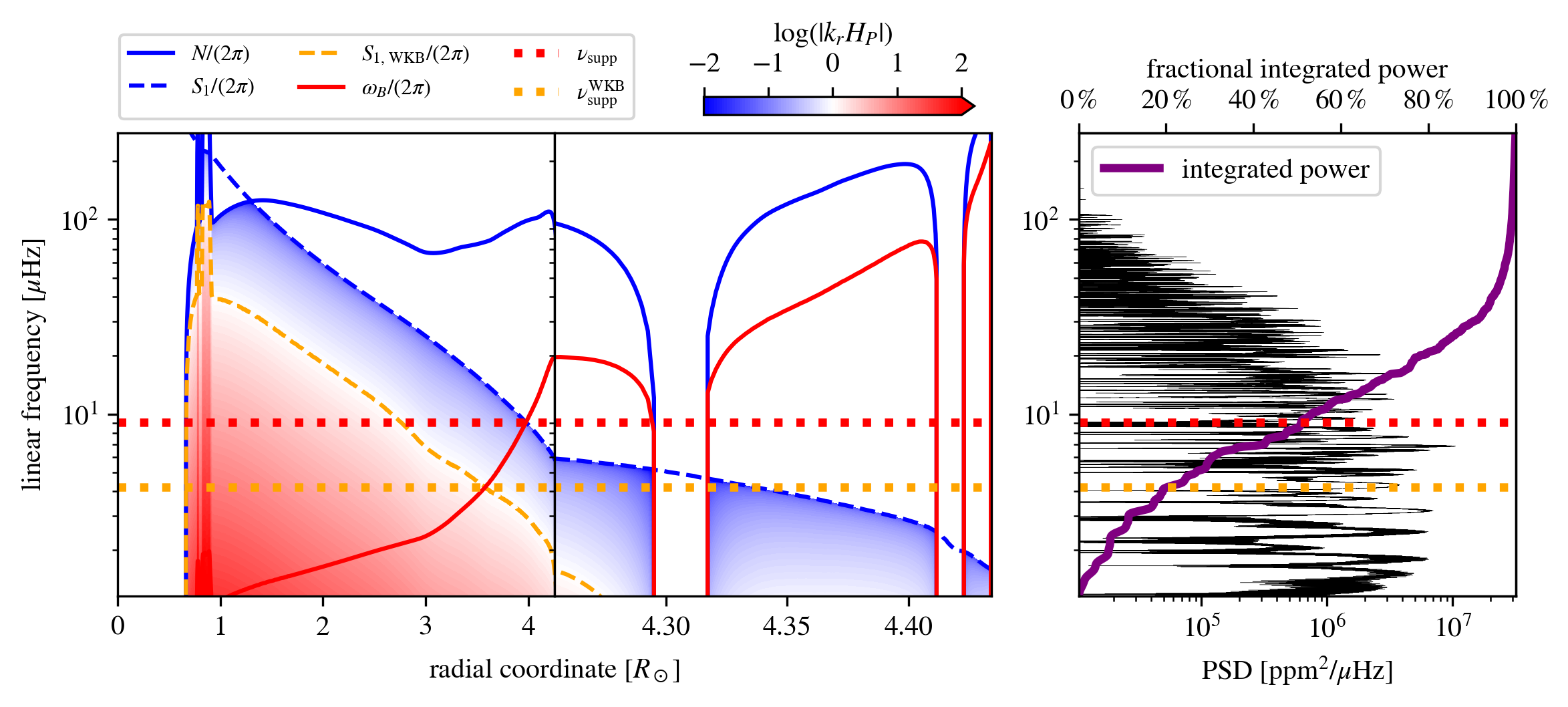}
    \caption{Propagation diagram color-coded by $\log(|k_rH_p|)$ for the HD 36485 model shown in Figure \ref{fig:propagation} (left panel), with the power spectrum of HD 36485's SLF variability reproduced in the right panel.
    Waves below $S_{1,\mathrm{WKB}}$ (dashed orange line) satisfy the WKB limit and experience magnetic suppression, while waves above this frequency experience a theoretically uncertain interaction with the magnetic field. 
    Excluding physically uncertain non-WKB regions reduces $\nusupp$ to a lower value $\nu_\mathrm{supp}^\mathrm{WKB}$ (dotted orange line).
    This reduces suppressed SLF power fraction somewhat, but not enough to restore internally driven IGWs as a viable origin for magnetic SLF variability, as a substantial SLF power below $\nu_\mathrm{supp}^\mathrm{WKB}$ are still observed.
    }
    \label{fig:wkb}
\end{figure*}

\citet{Fuller:2015:SuppressedDipole} demonstrate that IGWs become evanescent within magnetic suppression regions.
In the above analysis, we assume that any wave energy carried by IGWs instantaneously attenuates upon reaching the conditions for magnetic suppression.
However, a small fraction of power may still reach the stellar surface by tunneling through the evanescent region.

This fraction is very sensitive to both the local wavelength of the IGW and the thickness of the evanescent region.
We estimate the fraction of potentially visible wave energy by computing the tunneling coefficient for an outward IGW launched at a radius $r_0$:
\begin{equation}
\label{eq:tunnel}
T\simeq\exp\left(-2\int_{r_0}^{R}|\Im(k_r)|dr\right)\,,
\end{equation}
where $R$ is the radius of the star and $k_r$ is the local radial wavenumber.

We estimate $k_r$ by treating regions of qualitatively different wave propagation in a piecewise fashion.
These regions are classified in Figure \ref{fig:tunneling} for the best-fitting model for HD 36485.
In the blue, gray, and gold regions where $\omega>\omega_B$, dipole IGWs are not significantly affected by magnetic fields and therefore obey the classic gravito-acoustic (gp) dispersion relation \citep{Unno1989}:
\begin{equation}
\label{eq:kr_gp}
k_r^2=k_{r,\mathrm{gp}}^2=\frac{(\omega^2-S_1^2)(\omega^2-N^2)}{r^2\omega^2}\,.
\end{equation}
Waves remain propagating in the gravity-wave and pressure-wave regions where $\omega<\min(N,S_1)$ (blue) and $\omega>\max(N,S_1)$ (gray), respectively. As $\Im(k_r)=0$ in these propagating regimes, no evanescent attenuation occurs.
However, when $N<\omega<S_1$ or $S_1<\omega<N$ (denoted as ``gp-evanescent;'' gold regions in Figure \ref{fig:tunneling}), $k_r^2<0$ and wave energy attenuates as the waves become evanescent in those regions.

Waves with gravity-wave character ($\omega<\min(N,S_1)$) are also evanescent when $\omega<\omega_B$, i.e., the waves become magnetic-evanescent (red regions in Figure \ref{fig:tunneling}).
\citet{Fuller:2015:SuppressedDipole} estimates the local dispersion relation for magneto-gravity waves:
\begin{equation}
\label{eq:kr_mg}
    k_r^2=k_{r,\mathrm{mg}}^2=\frac{2\pi\rho\omega^2}{B_r^2}\left(1\pm\sqrt{1-\frac{\omega_B^4}{\omega^4}}\right)\,.
\end{equation}
The radial wavenumber $k_r$ is generally complex-valued in magnetic-evanescent regions where $\omega<\omega_B$ and has $\Im(k_r)\neq0$, which implies evanescent energy attenuation.

The right panel of Figure \ref{fig:tunneling} shows the energy tunneling coefficient $T$ as a function of frequency, where $r_0$ is taken to be the radius of the innermost propagating region.
For comparison, we also compute the tunneling coefficient using the conventional gp dispersion relation (Equation \ref{eq:kr_gp}), excluding magnetic contributions.
Without magnetic suppression, a non-negligible fraction of the wave energy may still reach the stellar surface below the suppression frequency $\nusupp$.
However, magnetic suppression drastically reduces this fraction by many orders of magnitude for low-frequency IGWs because the local evanescent depth in magnetic suppression regions is usually extremely small.
This validates our simplified assumption that IGWs which coincide with magnetic suppression regions cannot manifest as SLF variability.
This finding is typical of those found for all the other stars in our sample, as seen in Figures \ref{fig:HD_108} to \ref{fig:HD_57682} in Appendix \ref{app:full}.

\subsection{Non-WKB Effects on IGW Propagation}

Derivations of magnetic suppression have so far typically relied upon a local Wentzel--Kramers--Brillouin (WKB) analysis of IGWs, which assumes that the radial wavenumber $k_r$ locally satisfies $k_rH_P\gg 1$, where $H_P$ is the local pressure scale height \citep[see, e.g.,][]{Lecoanet2017,Rui2023,Rui:2024:magnetolaplace,David:2026:MagneticConversion}.
Under these conditions, gradients in background quantities are negligible, and the wave propagates in an energy-conserving fashion without reflection.
Although magnetic suppression likely occurs in some fashion under non-WKB conditions, it is not yet theoretically understood exactly how.

The left panel of Figure \ref{fig:wkb} locally evaluates $k_{r,\mathrm{gp}}H_P$ and overlays it in color upon the IGW propagation diagram for the best-fitting model for HD 36485.
As wave frequencies approach $S_1$, their wavenumbers drop to zero (cf. Equation \ref{eq:kr_gp}), and the waves leave the WKB limit.

To test the impact of using WKB-derived formulae in non-WKB regions, we attempt to exclude non-WKB regions from the magnetic suppression region for the purpose of evaluating $\nusupp$.
Shells with $k_{r,\mathrm{gp}}H_P=1$ define an effective frequency scale $S_{1,\mathrm{WKB}}$, where $\omega=S_{1,\mathrm{WKB}}$ estimates the outermost radius of gravity waves \textit{in the WKB regime} (Figure \ref{fig:wkb} for our HD 36485 model).
We then compute $\nu_\mathrm{supp}^\mathrm{WKB}$, given by the local $\omega_B/2\pi$ at the crossover radius between $S_{1,\mathrm{WKB}}$ and $\omega_B$. This effectively evaluates the suppression frequency over shells with $\omega<\min(N,S_{1,\mathrm{WKB}})$ rather than $\omega<\min(N,S_1)$.

For HD 36485, we obtain $\nu_\mathrm{supp}^\mathrm{WKB}\approx4.2\,\mu\mathrm{Hz}$ (dotted orange line in Figure \ref{fig:wkb}), a frequency scale that is still higher than $f_\mathrm{supp}^\mathrm{WKB}\approx20\%$ of HD 36485's SLF variability power (right panel of Figure \ref{fig:wkb}).
We repeat this test for all \numberofsystems systems in our sample (see Appendix \ref{app:full}).
Figure \ref{fig:suppressed_fractions} includes $f_\mathrm{supp}^\mathrm{WKB}$ alongside $f_\mathrm{supp}$ from our fiducial analysis.
While removing non-WKB regions necessarily reduces the suppression frequency, an IGW origin for SLF variability remains excluded for a handful of systems.
Our conclusions again hold when we repeat this test while instead assuming quadrupole ($\ell=2$) waves.
The results of these tests are reported in Table \ref{tab:results} for our entire sample.
In summary, our overall conclusions in this paper are robust against the exclusion of non-WKB propagation regions in which the underlying physics is uncertain.

\section{Conclusions and Outlook}
\label{sec:conclusion}

Internal gravity waves (IGWs) have been proposed to explain the origin of SLF variability in massive stars \citep{Bowman2019}.
However, many theoretical and observational studies in the recent literature have established that IGWs experience strong magnetic suppression when traveling through highly-magnetized layers.
In this work, we calculate the conditions for magnetic suppression to operate in models appropriate to \numberofsystems observed SLF variables with known magnetic fields.

We demonstrate that many systems in our sample exhibit significant SLF variability below frequencies $\nusupp$ where IGWs would be extremely magnetically damped.
Therefore, internally driven IGWs cannot explain the observed SLF power for these magnetic stars.
Given the similarity of the power spectrum of magnetic SLF variables to those of normal SLF variables, \textit{this challenges the IGW interpretation as a universal origin for SLF variability.}

Our simplified analysis is subject to a few theoretical caveats, we assess that none of them are likely to affect our ultimate conclusions. This work has, for example, only considered wave propagation and has set aside the question of wave excitation.
Our estimates of $\omega_B$ are derived from the approximate expression of Equation \ref{eqn:Bcrit} \citep[cf.][]{Fuller:2015:SuppressedDipole,Cantiello:2016:MagnetoAstero} which may change by an order-unity factor depending on the magnetic field geometry \citep{Lecoanet2017,Rui2023,DuerfeldtPedros2026,Rui:2026:magnetogeneral}.
Future work with more realistic treatments of field geometry and convective heating/radiative damping of IGWs will refine our first estimates of $\nusupp$.
Nevertheless, we expect that our overall conclusions are robust to these considerations.
Our findings are aligned with those of \citet{Anders2023}, who disfavor the IGW hypothesis by arguing that insufficient gravity-wave energy produced by core convection can reach the photosphere, even in the absence of magnetic fields.

Another proposed origin of SLF variability is the turbulent motions associated with surface convection in the stellar envelope \citep{Cantiello2021}.
Recent works modeling massive star envelopes with 3D radiation hydrodynamical simulations show that this mechanism can qualitatively produce SLF variability across a broad range of stellar spectral types \citep{Schultz2022,Ma:2026:BSGSimulation}.
The efficiency of near-surface convection is also expected to be modified by magnetic fields in a manner that warrants more exploration \citep[for similar concerns in the white dwarf literature, see, e.g.,][]{Tremblay:2015:MagneticWDs,Jewett:2025:UltramassivePulsators}. Future work should also consider the value of magnetic SLF variables for the assessment of the near-surface convection hypothesis.

In addition to SLF variability, massive stars also often exhibit ``macroturbulent'' spectroscopic line broadening, in excess to broadening caused by stellar rotation \citep{Howarth1997,Ryans2002,SimonDiaz2010,Serebriakova2023}.
Although its origin, too, remains uncertain, several studies have suggested that it may share a common physical origin with SLF variability, whether it arises from IGWs or surface convection or something else \citep{Aerts2009,Grassitelli2015,Serebriakova2024,Ma:2026:BSGSimulation}.
If so, magnetic SLF variables may be similarly diagnostic of the origin of macroturbulent line broadening.
For example, \citet{Sundqvist2013} reported suppressed macroturbulence in the highly magnetized star NGC 1624-2.
Holistic studies of SLF variability and macroturbulent line broadening in magnetic SLF variables may yield outsized insights into massive-star phenomenology.

\begin{acknowledgments}
We thank Lars Bildsten, Jim Fuller, and Eliot Quataert for helpful comments. L.M. is supported by the Gordon and Betty Moore Foundation Grant GBMF7392 at the Kavli Institute for Theoretical Physics (KITP), and the Lyman Spitzer, Jr. Postdoctoral Fellowship at Princeton University. N.Z.R. acknowledges support from the NASA Hubble Fellowship grant HST-HF2-51589.001-A awarded by the Space Telescope Science Institute, which is operated by the Association of Universities for Research in Astronomy, Inc., for NASA, under contract NAS5-26555. The collaborations between the authors have benefited from the Mitchel Postdoctoral Scholar Career Development Fund.
\end{acknowledgments}

\facilities{\textit{TESS} \citep{Ricker:2014:TESS}}

\software{\texttt{MESA} \citep{Paxton2011,Paxton2013,Paxton2015,Paxton2018,Paxton2019,Jermyn2023}}

\bibliography{bibliography}{}
\bibliographystyle{aasjournalv7}

\appendix

\section{Pre-whitening Processes}
\label{app:pre-whitening}

\begin{figure*}
    \centering
    \includegraphics[width=\textwidth]{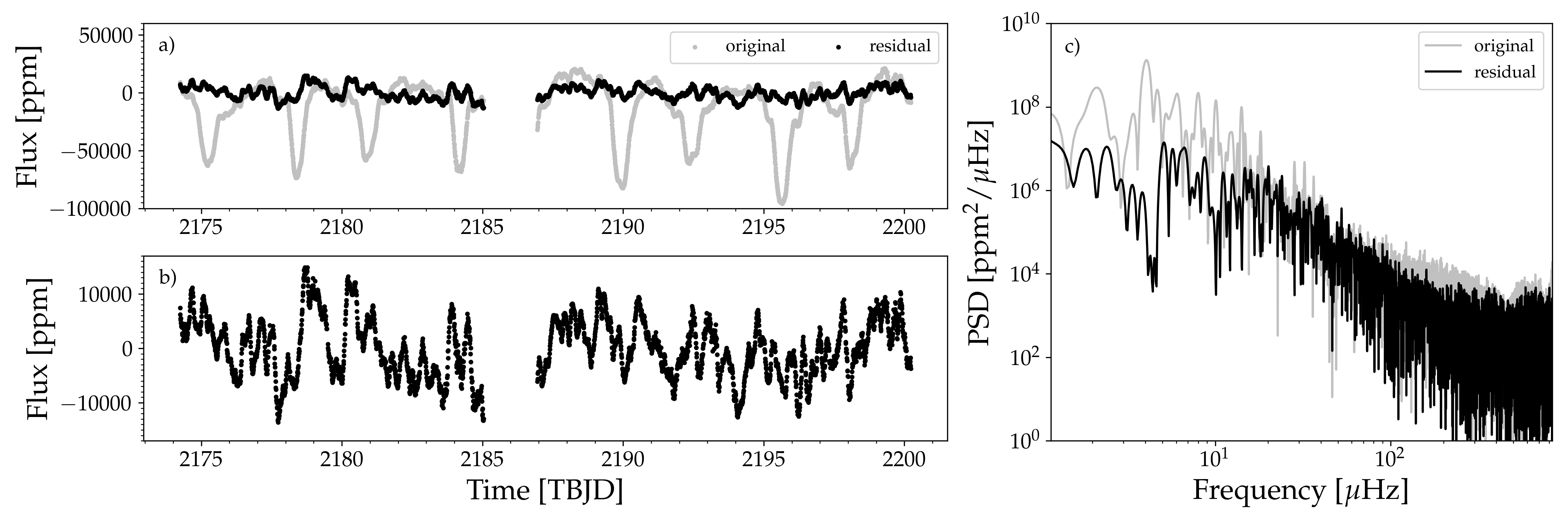}
    \caption{Example \textit{TESS} light curve and power spectral density of the eclipsing binary HD~36485 before (gray) and after (black) iterative prewhitening. The time is given in \textit{TESS} Barycentric Julian Date (TBJD). \emph{Panel a):} Both the original and residual light curves are shown. \emph{Panel b):} Only the residual light curve with the eclipses removed from prewhitening is shown. \emph{Panel c):} Corresponding power spectral density.
    }
    \label{fig:light_curve}
\end{figure*}

We perform pre-whitening on the normalized light curves for the system in our initial sample after removing contaminated targets. The prewhitening is done separately for consecutive \textit{TESS} sectors for each target. At each step of the prewhitening, we calculate the Lomb--Scargle periodogram \citep{Lomb1976,Scargle1982} of the residual light curve and identify the highest signal-to-noise (S/N) peak. The noise is estimated in the periodogram by calculating a running median of the periodogram with a window width of $2\,{\rm d}^{-1}$. The initial periodogram is calculated with an oversampling factor of 10, such that the sampling in the frequency space is given as $\nu_s = \nu_R/10$ where $\nu_R = 1/T$ is the Rayleigh limit and $T$ is the length of the time series. Once the highest S/N peak has been identified, the periodogram is recalculated in a small frequency interval $[\nu_{\rm peak} - \nu_R, \nu_{\rm peak} + \nu_R]$ with 1000 equally space frequency bins around the frequency $\nu_{\rm peak}$ of the highest S/N peak. The amplitude and frequency of the highest S/N peak is then redetermined for this increased sampling in frequency space and used when fitting a model $F_{\rm model}(t)$ consisting of a sum of sinusoids to the original light curve:

\begin{equation}
    F_{\rm model}(t) = C + \sum^N_i A_i \sin\left(2 \pi \left[\nu_i t + \phi_i \right] \right).
    \label{eq:sin_model}
\end{equation}

Here $C$ is a constant, and $A_i$, $\nu_i$, and $\phi_i$ are the amplitudes, frequencies, and phases of the $i = 1, \dots, N$ signals to be fit to the original light curve $F_{\rm original}(t)$ using \texttt{lmfit} \citep{lmfit}. During the fit the frequencies are held fixed while the amplitude of the most recent signal $A_N$ is optimized together with all the phases $\phi_i$. Once a best fit model for the current iteration has been obtained, the residual light curve $F_{\rm residual} (t) = F_{\rm original}(t) - F_{\rm model}(t)$ is calculated, and the next highest S/N peak in the resulting periodogram is identified. Following \citet{Pedersen2025}, the prewhitening is stopped once the highest S/N signal in the residual periodogram drops below $S/N = 4$. Figure~\ref{fig:light_curve} shows an example light curve and corresponding power spectral density before (gray) and after (black) iterative prewhitening.

\section{Summary for All \numberofsystems Systems}

In this appendix, we present the summary of observations and results for all 17 SLF variables in our sample. Table \ref{tab:observed} summarizes the observed stellar parameters (luminosity, effective temperature, and surface gravity) taken from \cite{Shen2023}, the measured constraints on the polar magnetic field $B_\mathrm{p}$ taken from \cite{Petit2013}, and the \textit{TESS} observing sectors in this work. Table \ref{tab:results} summarizes the best-fitted stellar model parameters, the assumed radial magnetic field strength $B_r$, and our calculated suppression frequency $\nu_\mathrm{supp}$ and suppressed SLF power fraction $f_\mathrm{supp}$.

\input{system_table}

\input{results}

\section{Observations and Results for the Full Sample}
\label{app:full}

Figures \ref{fig:HD_108} through \ref{fig:HD_57682} show the calculations of magnetic suppression and the observed SLF variability for the full 17 stars in our sample. The left two panels show similar propagation diagrams and observed periodograms for each star, as in Figure \ref{fig:wkb}. The right panels show the wave energy tunneling coefficients similar to the right panel of Figure \ref{fig:tunneling}.

\clearpage
\input{supplementary}

\end{CJK*}
\end{document}

%% file: system_table.tex
\begin{table*}[htbp]
\centering
\begin{tabularx}{0.95\textwidth}{llccccl}
\hline\\[-3mm]
star name & TIC & $\log(L/L_\odot)$ & $\log(T_\mathrm{eff}/\mathrm{K})$ & $\log g\;[\mathrm{cm\,s}^{-2}]$ & $B_\mathrm{p}\;[\mathrm{kG}]$ & \textit{TESS} FFI sectors observed\\[1mm]
\hline

HD 108 & 406893156 & $5.7\pm0.1$ & $4.54\pm0.02$ & $3.5\pm0.2$ & $>0.5$ & 17, 18, 57, 58, 77, 78, 84, 85 \\
HD 125823 & 167480388 & $3.2\pm0.2$ & $4.28\pm0.05$ & $4.14\pm0.12$ & $>1.3$ & 11, 38, 65, 102 \\
HD 127381 & 457686485 & $3.76\pm0.06$ & $4.36\pm0.02$ & $4.02\pm0.1$ & $0.5$ & 11, 38, 65, 101 \\
HD 130807 & 455309397 & $2.7\pm0.2$ & $4.23\pm0.03$ & $4.25\pm0.1$ & $>2$ & 11, 38, 65, 102, 103 \\
HD 136504 & 147226597 & $3.7\pm0.2$ & $4.31\pm0.01$ & $3.97\pm0.15$ & $>0.6$ & 38, 65, 102, 103 \\
HD 149277 & 229331994 & $3.5\pm0.2$ & $4.3\pm0.04$ & $3.75\pm0.15$ & $>4.7$ & 12, 39, 66, 93 \\
HD 191612 & 378273410 & $5.4\pm0.2$ & $4.54\pm0.01$ & $3.5\pm1$ & $2.5$ & 14, 15, 41, 54, 55, 75, 81, 82 \\
HD 200775 & 387813185 & $4\pm0.3$ & $4.25\pm0.05$ & $3.4\pm0.2$ & $1$ & 16, 18, 24, 25, 52, 56--58, 76--78, 83--86 \\
HD 205021 & 321818578 & $4.3\pm0.1$ & $4.4\pm0.02$ & $3.8\pm0.15$ & $0.36$ & 16--18, 24, 25, 52, 57--59, 77--79, 84--86 \\
HD 25558 & 407615454 & $2.8\pm0.3$ & $4.23\pm0.02$ & $4.2\pm0.2$ & $>0.15$ & 5, 32 \\
HD 36485 & 50743458 & $3.5\pm0.1$ & $4.3\pm0.04$ & $4\pm0.1$ & $10$ & 6, 32, 98 \\
HD 37058 & 427393350 & $2.9\pm0.1$ & $4.27\pm0.01$ & $4.17\pm0.07$ & $>3$ & 6, 32, 98 \\
HD 37742 & 11360636 & $5.6\pm0.1$ & $4.46\pm0.01$ & $3.2\pm0.1$ & $0.06$ & 6, 98 \\
HD 46328 & 47763235 & $4.5\pm0.1$ & $4.43\pm0.02$ & $3.78\pm0.07$ & $>1.5$ & 6, 33, 87, 107 \\
HD 47129 & 220197273 & $5.09\pm0.04$ & $4.52\pm0.03$ & $4.1\pm0.1$ & $>2.8$ & 6, 87 \\
HD 55522 & 65403216 & $3\pm0.2$ & $4.24\pm0.01$ & $3.95\pm0.06$ & $>2.6$ & 7, 33, 34, 61, 87, 88 \\
HD 57682 & 187458882 & $4.8\pm0.2$ & $4.53\pm0.01$ & $4\pm0.2$ & $1.7$ & 7, 34 \\

\hline
\end{tabularx}
\caption{Observed properties of magnetic stars in our analysis. Measurements for luminosity, effective temperature, and surface gravity are taken from \cite{Shen2023}. The polar magnetic field measurements/constraints are taken from \cite{Petit2013}.}
\label{tab:observed}
\end{table*}

%% file: results.tex
\begin{table*}[htbp]
\centering
\begin{tabularx}{\textwidth}{l ccccc cccc cccc}
\hline

& \multicolumn{5}{c}{model parameters} & \multicolumn{4}{c}{$\ell=1$, dipole} & \multicolumn{4}{c}{$\ell=2$, quadrupole} \\
\cmidrule(lr){2-6}\cmidrule(lr){7-10}\cmidrule(lr){11-14}

star name & mass & $\log(L)$ & $\log(T_\mathrm{eff})$ & $\log g$ & $B_r$& $\nu_\mathrm{supp}$& $f_\mathrm{supp}$ & $\nu_\mathrm{supp}^\mathrm{WKB}$& $f_\mathrm{supp}^\mathrm{WKB}$ & $\nu_\mathrm{supp}$& $f_\mathrm{supp}$ & $\nu_\mathrm{supp}^\mathrm{WKB}$ &$f_\mathrm{supp}^\mathrm{WKB}$ \\

\quad[unit] & [$M_\odot$] & [$L_\odot$] & [$\mathrm{K}$] & $[\mathrm{cm\,s}^{-2}]$ &[kG] & [$\mu\mathrm{Hz}$] &  & [$\mu\mathrm{Hz}$] &  & [$\mu\mathrm{Hz}$] & & [$\mu\mathrm{Hz}$]  \\[2mm]
\hline

HD 108 & $46$ & $5.71$ & $4.54$ & $3.51$ & $0.5$ & 3.6 & 30\% & 2.6 & 21\% & 4.7 & 38\% & 4.5 & 37\% \\
HD 125823 & $6.25$ & $3.19$ & $4.28$ & $4.13$ & $1.3$ & 7.7 & 26\% & 3.7 & 9\% & 11.1 & 43\% & 5.8 & 15\% \\
HD 127381 & $8.75$ & $3.76$ & $4.36$ & $4.01$ & $0.5$ & 4.5 & 14\% & 2.5 & 5\% & 6.9 & 32\% & 3.9 & 13\% \\
HD 130807 & $5$ & $2.76$ & $4.24$ & $4.28$ & $2$ & 10.6 & 56\% & 5.1 & 21\% & 16.4 & 78\% & 8.0 & 43\% \\
HD 136504 & $7.5$ & $3.60$ & $4.31$ & $3.91$ & $0.6$ & 4.2 & 36\% & 2.4 & 13\% & 6.6 & 57\% & 3.7 & 34\% \\
HD 149277 & $7.25$ & $3.60$ & $4.29$ & $3.80$ & $4.7$ & 6.3 & 45\% & 3.0 & 22\% & 10.0 & 70\% & 4.8 & 35\% \\
HD 191612 & $33$ & $5.42$ & $4.54$ & $3.65$ & $2.5$ & 4.0 & 26\% & 2.1 & 9\% & 7.0 & 42\% & 3.6 & 25\% \\
HD 200775 & $8.5$ & $3.98$ & $4.26$ & $3.37$ & $1$ & 2.5 & 7\% & 1.4 & 2\% & 3.4 & 12\% & 2.2 & 6\% \\
HD 205021 & $12$ & $4.31$ & $4.41$ & $3.79$ & $0.36$ & 3.0 & 7\% & 1.7 & 2\% & 4.6 & 16\% & 2.9 & 7\% \\
HD 25558 & $5$ & $2.81$ & $4.23$ & $4.20$ & $0.15$ & 5.8 & 56\% & 2.8 & 28\% & 9.2 & 68\% & 4.1 & 39\% \\
HD 36485 & $7.25$ & $3.49$ & $4.31$ & $4.00$ & $10$ & 9.1 & 51\% & 4.2 & 20\% & 14.3 & 67\% & 6.6 & 36\% \\
HD 37058 & $5.75$ & $2.99$ & $4.27$ & $4.23$ & $3$ & 10.3 & 71\% & 4.9 & 36\% & 16.3 & 86\% & 7.8 & 58\% \\
HD 37742 & $38.5$ & $5.62$ & $4.46$ & $3.19$ & $0.06$ & $<\nu_\mathrm{min}$ & N/A &$<\nu_\mathrm{min}$ & N/A &1.5 & 2\% & $<\nu_\mathrm{min}$ & N/A \\
HD 46328 & $14$ & $4.51$ & $4.44$ & $3.77$ & $1.5$ & 4.4 & 23\% & 1.9 & 5\% & 7.2 & 47\% & 3.2 & 17\% \\
HD 47129 & $25.5$ & $5.05$ & $4.56$ & $3.97$ & $2.8$ & 5.0 & 48\% & 2.0 & 20\% & 8.2 & 66\% & 3.4 & 32\% \\
HD 55522 & $5.5$ & $3.10$ & $4.23$ & $3.97$ & $2.6$ & 7.5 & 46\% & 3.7 & 29\% & 11.8 & 53\% & 5.8 & 38\% \\
HD 57682 & $21$ & $4.82$ & $4.53$ & $4.02$ & $1.7$ & 5.2 & 26\% & 2.1 & 3\% & 8.5 & 36\% & 3.5 & 10\% \\

\hline
\end{tabularx}
\caption{The \texttt{MESA} model parameters, assumed minimal internal magnetic field $B_r$, and the calculated suppression frequencies and power fraction for dipole ($\ell=1$) and quadrupole ($\ell=2$) gravity waves in our analysis. Results excluding non-WKB waves are denoted by $\nu_\mathrm{supp}^\mathrm{WKB}$ and $f_\mathrm{supp}^\mathrm{WKB}$. For most systems, a substantial fraction of SLF power is expected to be suppressed if they are from IGWs.}
\label{tab:results}
\end{table*}

%% file: supplementary.tex
\begin{figure*}
\centering
\includegraphics[width=\textwidth]{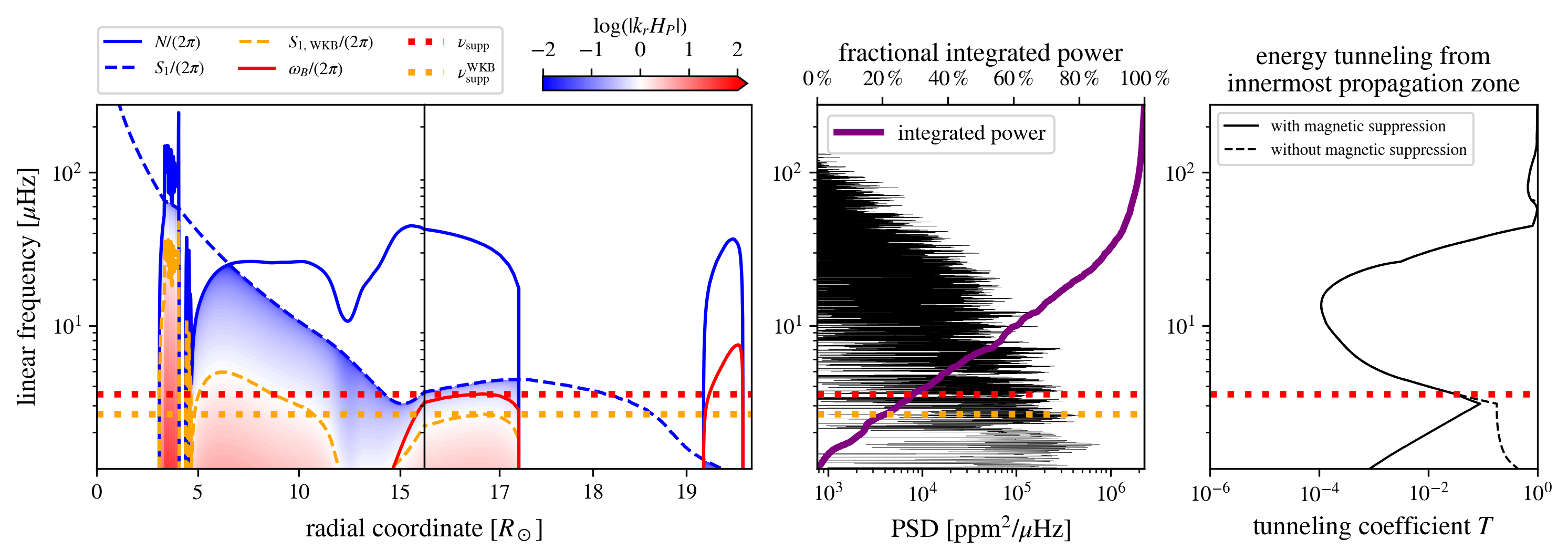}
\caption{
Summary of results and observations for HD 108. 
}
\label{fig:HD_108}
\end{figure*}
 
\begin{figure*}
\centering
\includegraphics[width=\textwidth]{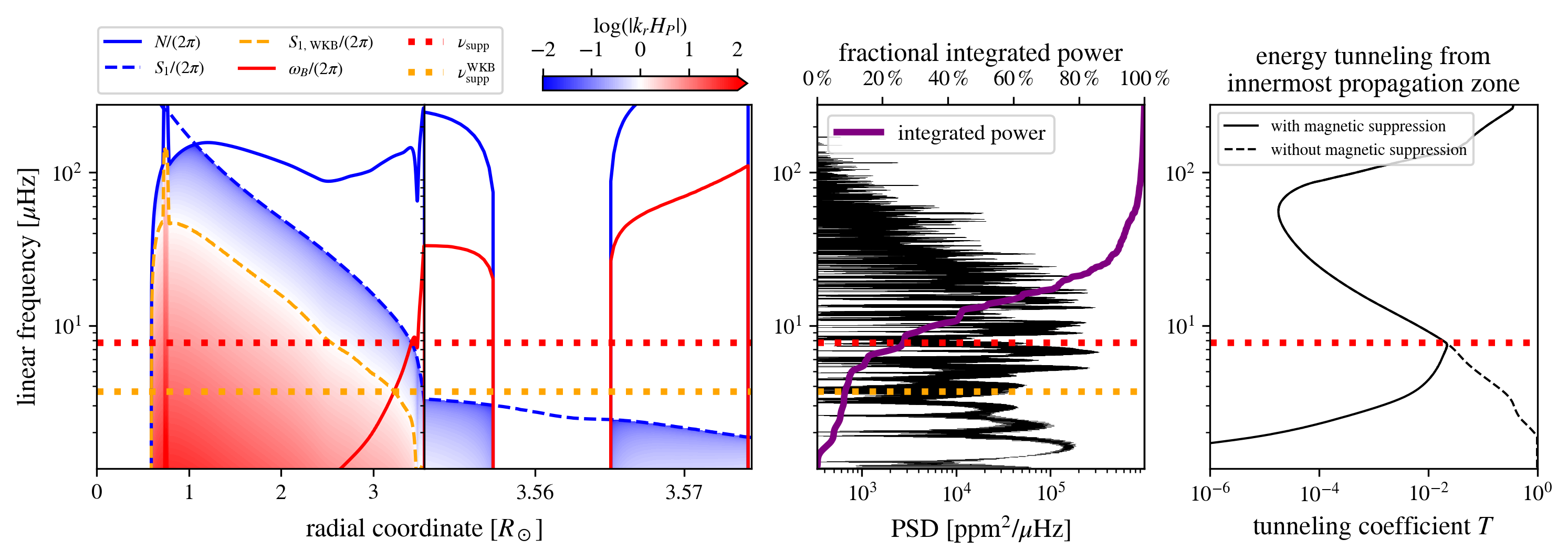}
\caption{
Summary of results and observations for HD 125823. 
}
\label{fig:HD_125823}
\end{figure*}
 
\begin{figure*}
\centering
\includegraphics[width=\textwidth]{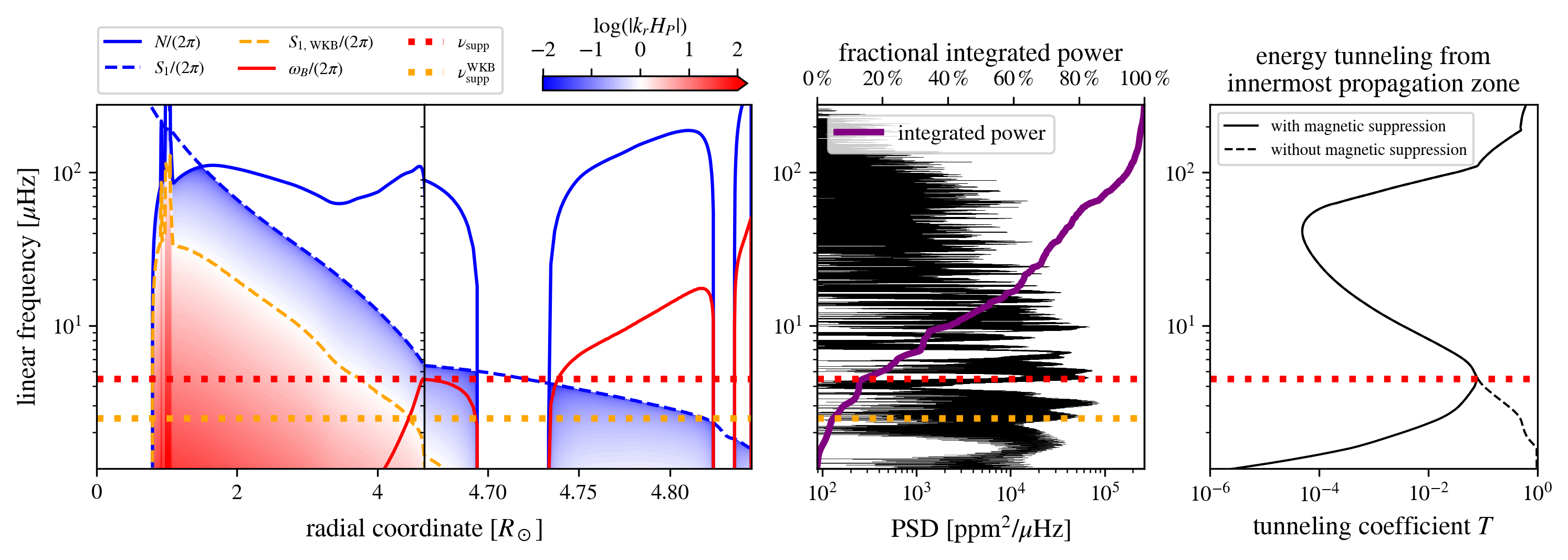}
\caption{
Summary of results and observations for HD 127381. 
}
\label{fig:HD_127381}
\end{figure*}
 
\begin{figure*}
\centering
\includegraphics[width=\textwidth]{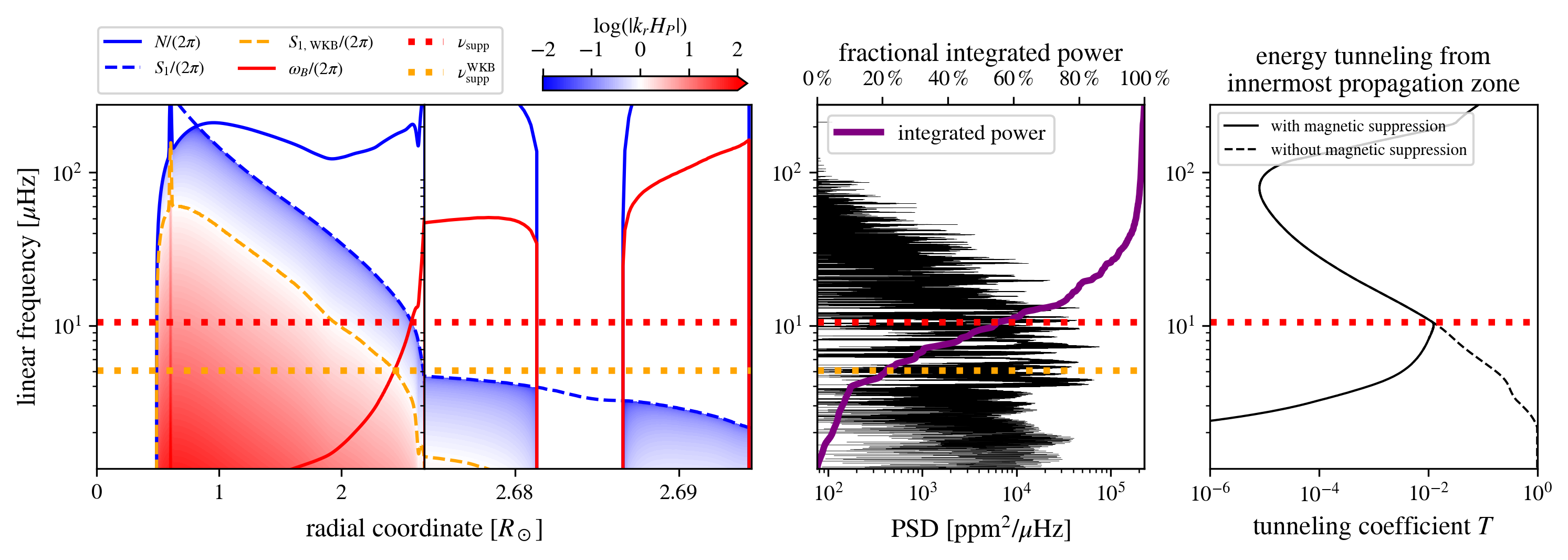}
\caption{
Summary of results and observations for HD 130807. 
}
\label{fig:HD_130807}
\end{figure*}
 
\begin{figure*}
\centering
\includegraphics[width=\textwidth]{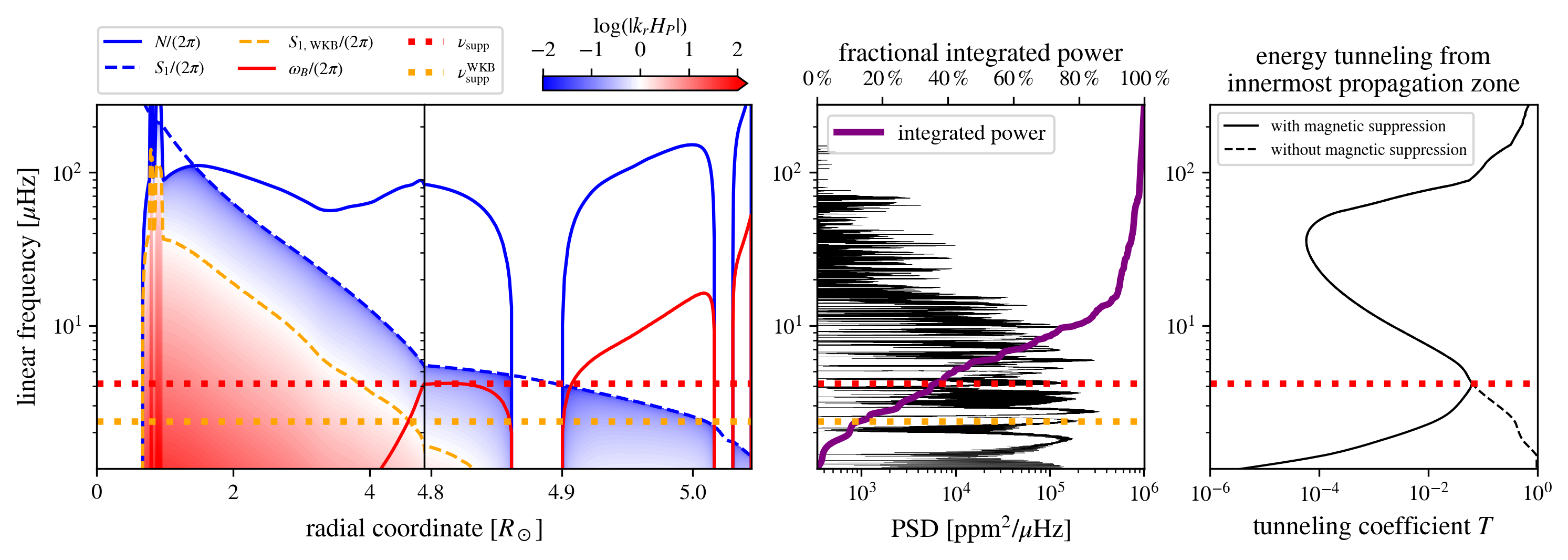}
\caption{
Summary of results and observations for HD 136504. 
}
\label{fig:HD_136504}
\end{figure*}
 
\begin{figure*}
\centering
\includegraphics[width=\textwidth]{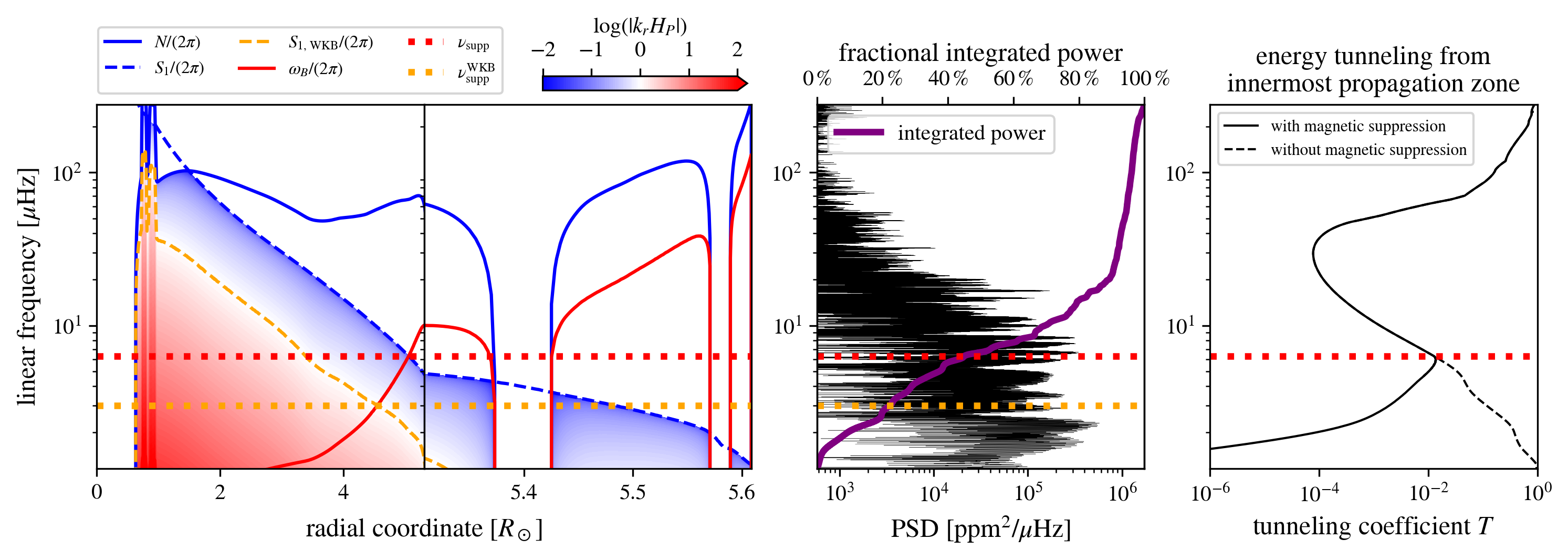}
\caption{
Summary of results and observations for HD 149277. 
}
\label{fig:HD_149277}
\end{figure*}
 
\begin{figure*}
\centering
\includegraphics[width=\textwidth]{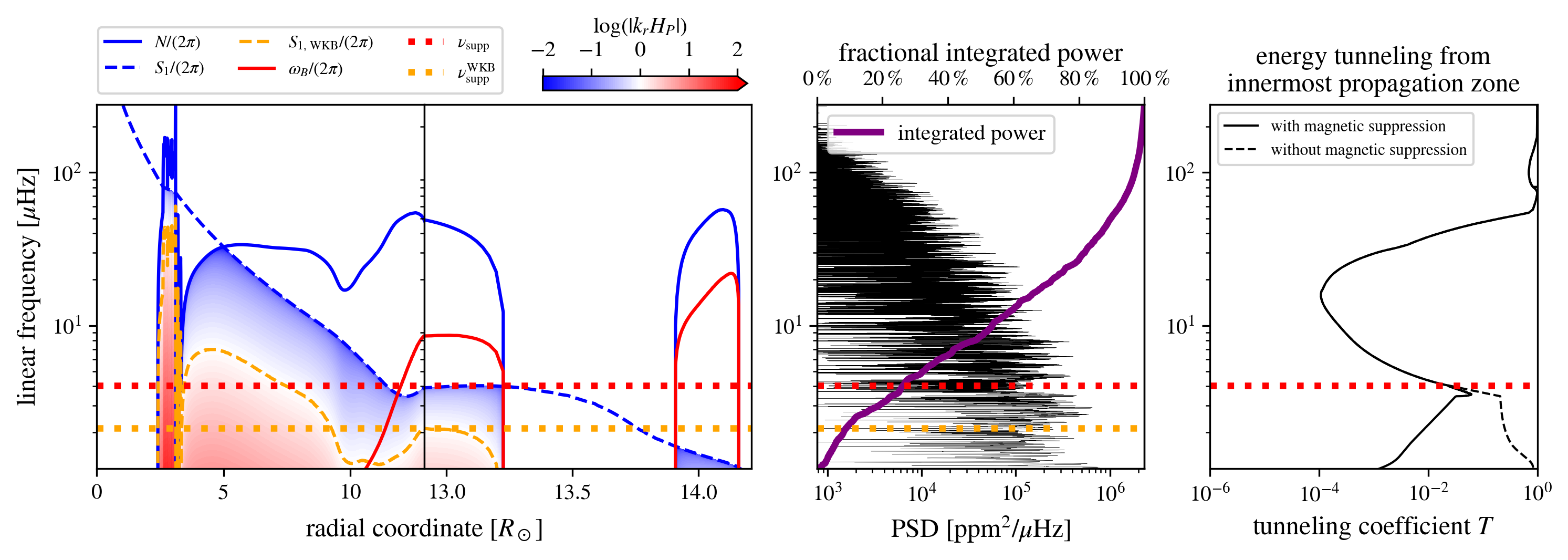}
\caption{
Summary of results and observations for HD 191612. 
}
\label{fig:HD_191612}
\end{figure*}
 
\begin{figure*}
\centering
\includegraphics[width=\textwidth]{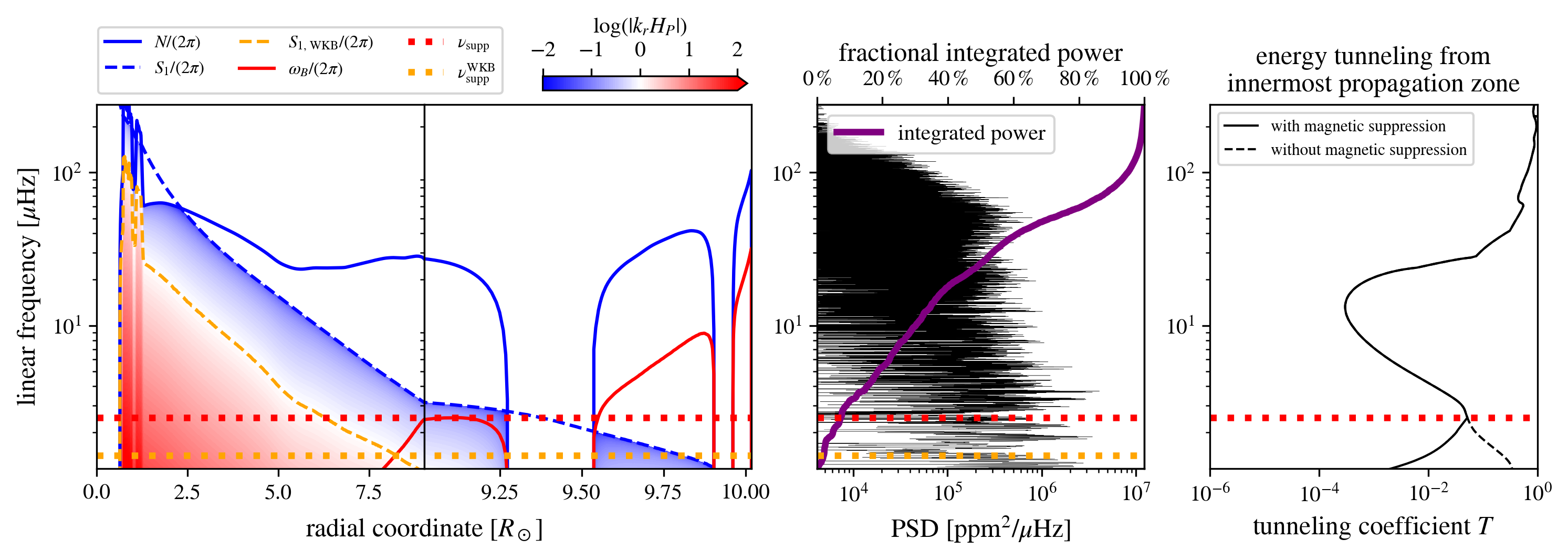}
\caption{
Summary of results and observations for HD 200775. 
}
\label{fig:HD_200775}
\end{figure*}
 
\begin{figure*}
\centering
\includegraphics[width=\textwidth]{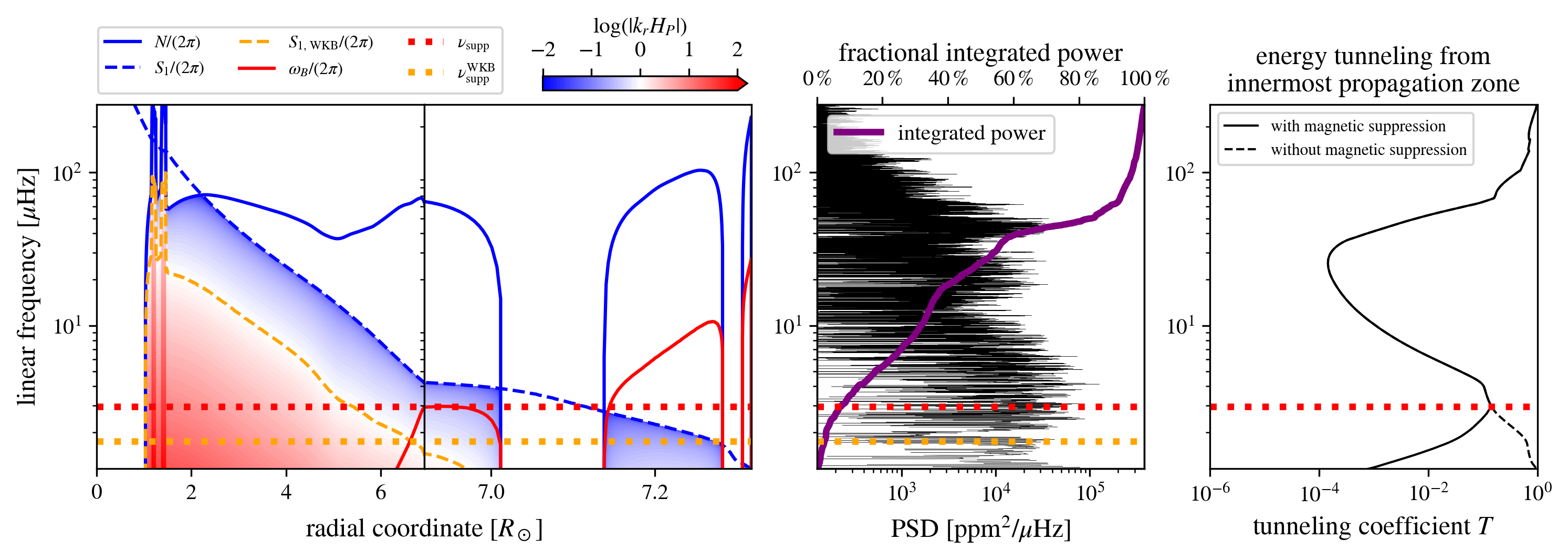}
\caption{
Summary of results and observations for HD 205021. 
}
\label{fig:HD_205021}
\end{figure*}
 
\begin{figure*}
\centering
\includegraphics[width=\textwidth]{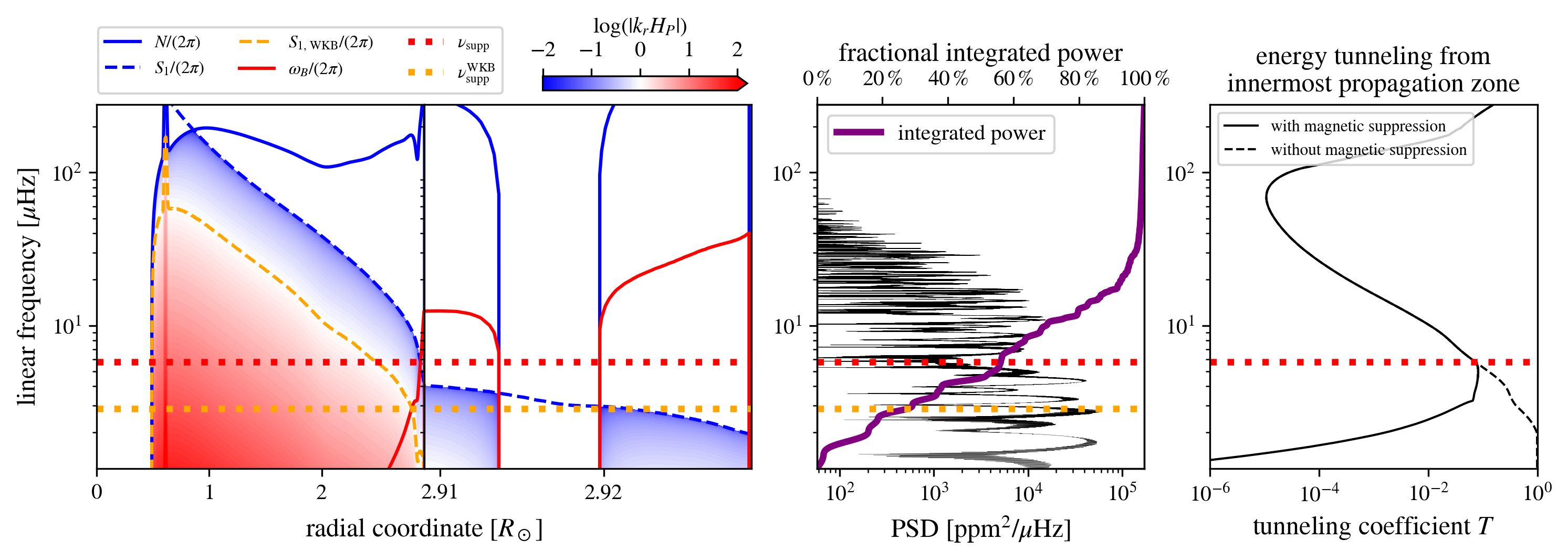}
\caption{
Summary of results and observations for HD 25558. 
}
\label{fig:HD_25558}
\end{figure*}
 
\begin{figure*}
\centering
\includegraphics[width=\textwidth]{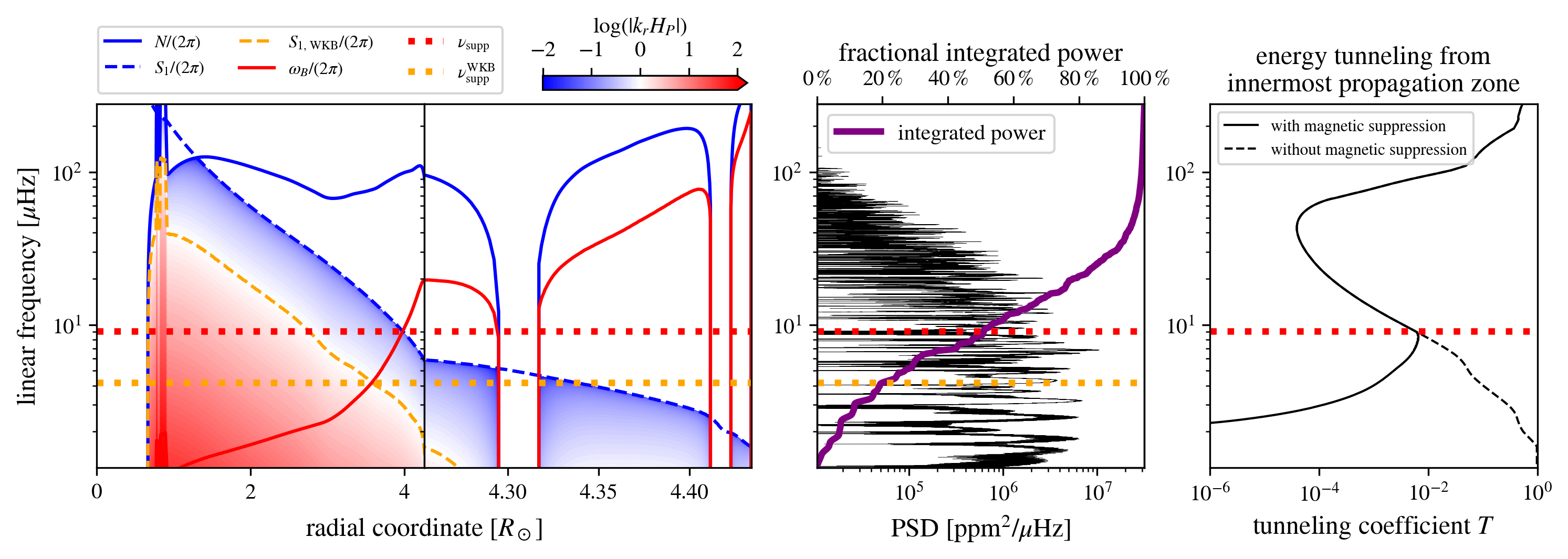}
\caption{
Summary of results and observations for HD 36485. 
}
\label{fig:HD_36485}
\end{figure*}
 
\begin{figure*}
\centering
\includegraphics[width=\textwidth]{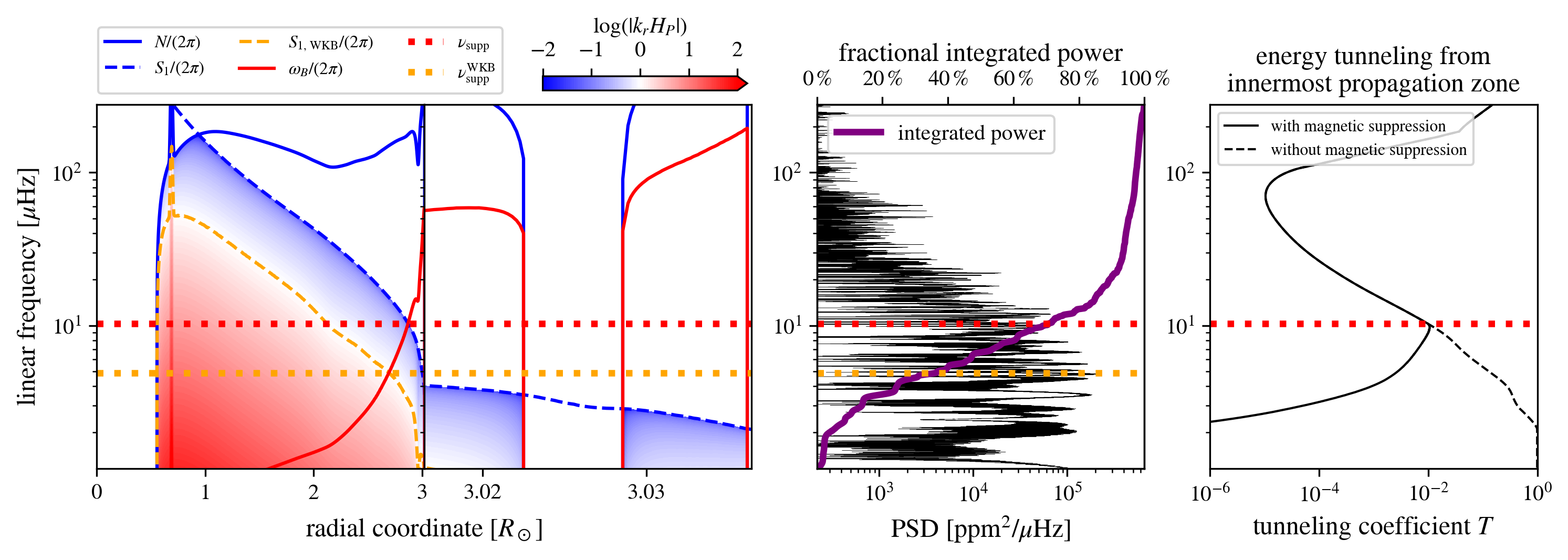}
\caption{
Summary of results and observations for HD 37058. 
}
\label{fig:HD_37058}
\end{figure*}
 
\begin{figure*}
\centering
\includegraphics[width=\textwidth]{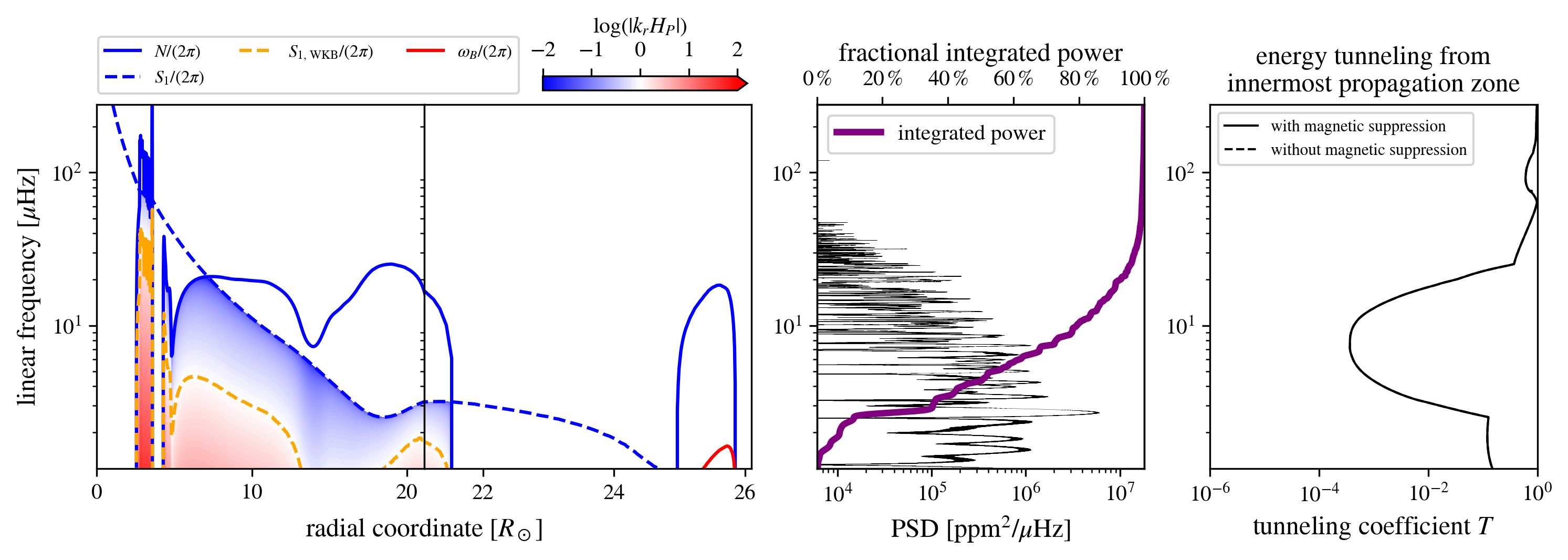}
\caption{
Summary of results and observations for HD 37742. Magnetic field in this star is not strong enough to suppress SLF power above the lowest frequency of the PSD ($\nu_\mathrm{min}=1.157\,\mu\mathrm{Hz}$).
}
\label{fig:HD_37742}
\end{figure*}
 
\begin{figure*}
\centering
\includegraphics[width=\textwidth]{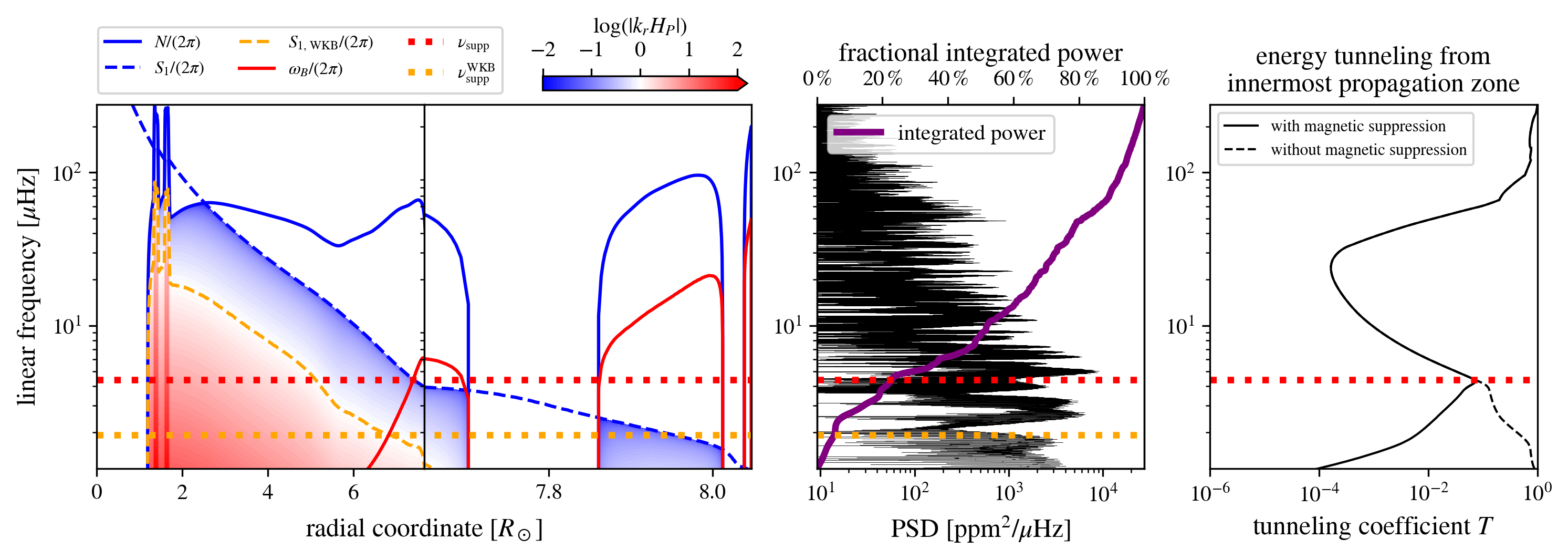}
\caption{
Summary of results and observations for HD 46328. 
}
\label{fig:HD_46328}
\end{figure*}
 
\begin{figure*}
\centering
\includegraphics[width=\textwidth]{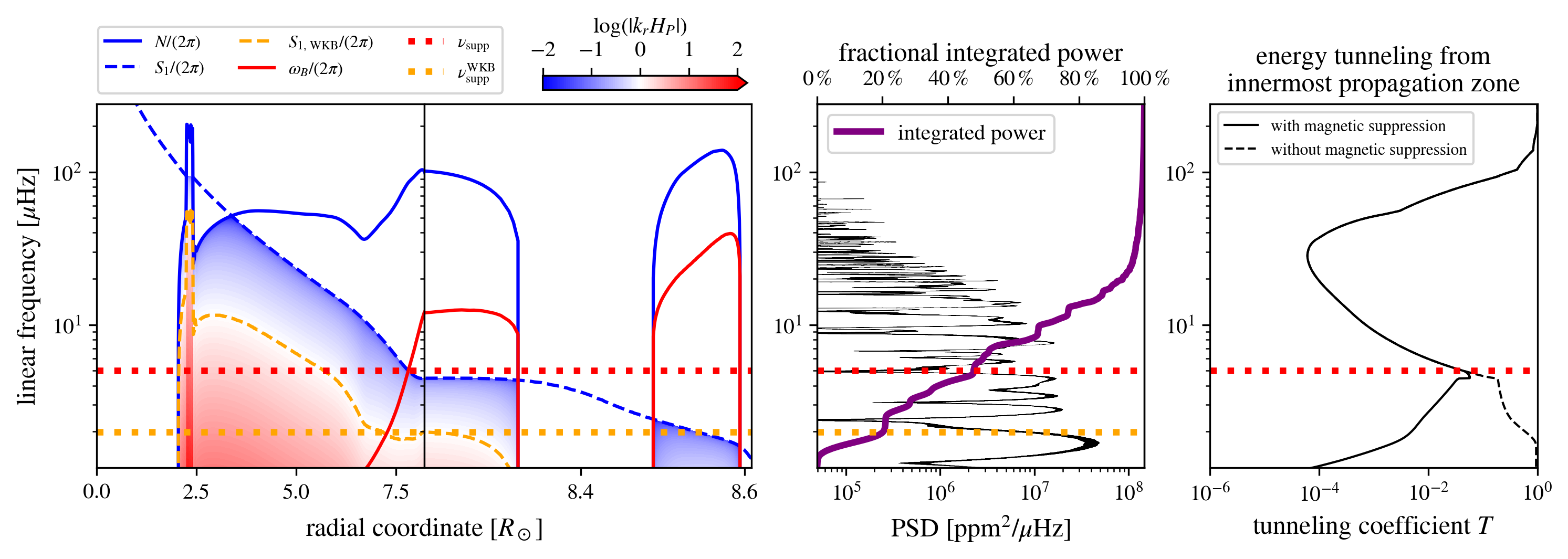}
\caption{
Summary of results and observations for HD 47129. 
}
\label{fig:HD_47129}
\end{figure*}
 
\begin{figure*}
\centering
\includegraphics[width=\textwidth]{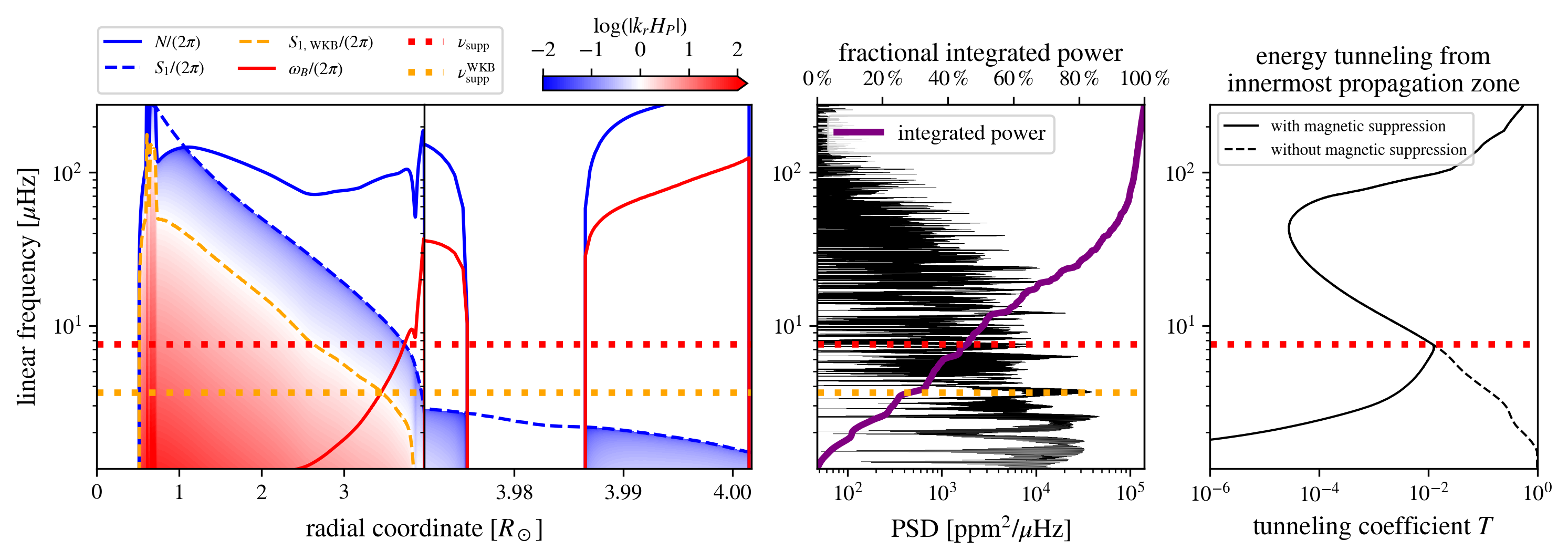}
\caption{
Summary of results and observations for HD 55522. 
}
\label{fig:HD_55522}
\end{figure*}
 
\begin{figure*}
\centering
\includegraphics[width=\textwidth]{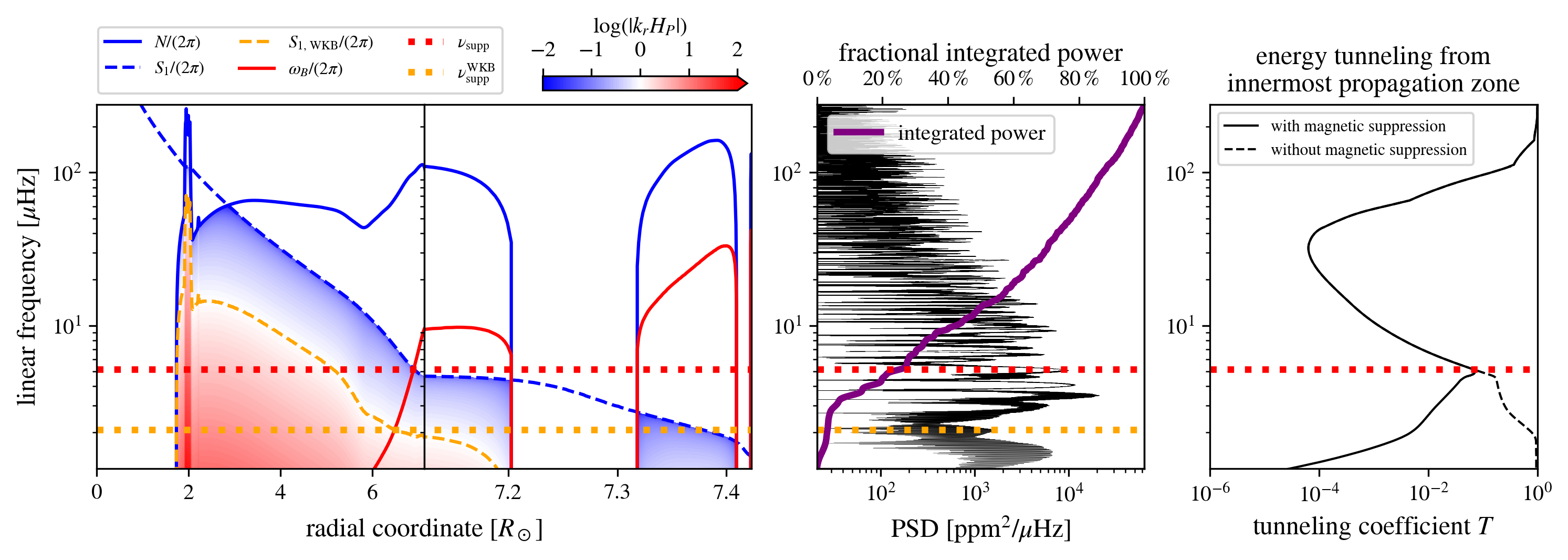}
\caption{
Summary of results and observations for HD 57682. 
}
\label{fig:HD_57682}
\end{figure*}